\documentclass{article}

\usepackage[preprint]{neurips_2026}
\usepackage{wrapfig}
\usepackage[utf8]{inputenc}
\usepackage[T1]{fontenc}
\usepackage{url}
\usepackage{booktabs}
\usepackage{amsfonts}
\usepackage{amsmath}
\usepackage{nicefrac}
\usepackage{microtype}
\usepackage{xcolor}
\usepackage{graphicx}
\usepackage{multirow}
\usepackage{subcaption}
\usepackage{tabularx}
\usepackage{array}
\usepackage{natbib}
\usepackage{listings}
\newcolumntype{Y}{>{\centering\arraybackslash}X}
\usepackage{tikz}
\usetikzlibrary{positioning,arrows.meta}
\usepackage{xspace}
\PassOptionsToPackage{hyphens}{url}
\usepackage[
    hidelinks
]{hyperref}
\newcommand{\Name}{CacheCraft\xspace}

\newcommand{\missingmeasurement}[1]{\textit{not measured}}

\title{Discovering KV Cache Eviction Policies\\via LLM-Guided Program Evolution}

\author{
Pratik Poudel \\
FIU \\
Miami, FL, USA \\
\texttt{ppoud001@fiu.edu}
\And
Yanzhao Wu \\
FIU \\
Miami, FL, USA \\
\texttt{yawu@fiu.edu}
\And
Sumit Jha \\
FIU \\
Gainesville, FL, USA \\
\texttt{sumit.jha@ufl.edu}
\And
Jason Liu \\
FIU \\
Miami, FL, USA \\
\texttt{liux@fiu.edu}
}

\begin{document}

\maketitle


\begin{abstract}
KV cache compression is critical for long-context inference, yet effective eviction policies remain difficult to design: existing prefill-stage methods often rely on hand-crafted salience heuristics that can be brittle across models, context lengths, and compression ratios. We present \textbf{\Name}, a program-evolution methodology for automatically discovering KV cache eviction policies using an LLM-guided code-evolution engine. \Name discovers \textbf{FRC} (Feature-Rich Compression), a fixed-weight three-signal scorer that combines local attention received, neighborhood attention density, and KV-head maximum salience with chunk-level top-$k$ selection. Without per-model retuning, FRC ranks first among the evaluated single-pass KVPress baselines at every RULER 4k/8k cell with $r \geq 0.75$ across Llama-3.1-8B-Instruct and Qwen3-8B (12 of 20 grid cells), gaining $+15.4$ points on Llama-4k and $+13.9$ points on Qwen-8k at $88\%$ compression. A scorer-versus-structure decomposition shows that the scoring family, not chunk selection, is the load-bearing design choice: incorporating the scorer contributes $+67.2$ RULER points, while improving chunk structure contributes only $\approx 0.1$. Beyond FRC itself, \Name provides a transferable recipe for automated eviction-policy discovery: a compact policy interface, a cascade evaluator with strict output invariants, and a diagnostic loop that treats search plateaus and reward-hacking failures as evidence for reformulating the editable interface.
\end{abstract}

\section{Introduction}\label{sec:intro}

Long-context language-model inference is increasingly constrained by the key-value (KV) cache. During prefill, the model materializes keys and values for every prompt token; during decoding, this state is repeatedly streamed from GPU memory. For long prompts, the cache can rival or exceed the memory footprint of the model weights themselves~\citep{pope2023efficiently,kwon2023vllm}. Prefill-stage eviction addresses this bottleneck by compressing the cache after the initial forward pass and before the first generated token.

Most existing prefill-stage methods rely on a single hand-designed salience statistic. SnapKV~\citep{li2024snapkv} pools attention over a recent observation window; ChunkKV~\citep{liu2025chunkkv} aggregates SnapKV scores into fixed chunks; ExpectedAttention~\citep{devoto2025expected} uses key norms as a query-independent proxy; and Finch~\citep{corallo2024finch} is a query-aware fusion baseline. This single-statistic design is brittle in the aggressive-compression regime that matters most for long-context throughput. In our evaluation, the strongest baseline reorders across models: ChunkKV leads on Llama-3.1-8B-Instruct at aggressive ratios, while ExpectedAttention is strongest on Qwen3-8B at 4k in some settings. Accuracy also collapses as the retained cache shrinks. These failures motivate a training-free eviction scorer that combines complementary salience signals and transfers across models and context lengths without per-model retuning.

Rather than hand-design another attention statistic, we treat eviction-policy design as a program-search problem. We present \textbf{\Name}, a program-evolution methodology that evolves prefill-stage KV eviction rules as editable program text. \Name exposes a compact KVPress~\citep{kvpress} policy module with two entry points, \texttt{score\_tokens} and \texttt{select\_tokens\_to\_keep}; gates candidates through a three-stage cascade evaluator with strict output invariants; and drives mutations with OpenEvolve~\citep{openevolve}, an LLM-guided code-evolution engine in the lineage of FunSearch~\citep{romera2024mathematical} and AlphaEvolve~\citep{novikov2025alphaevolve}. The methodological commitment is diagnostic: when search plateaus or reward hacking appears, we treat the failure as evidence that the editable interface, seed family, or evaluator contract is mis-scoped, and reformulate that component rather than merely spending more iterations.

\begin{figure}[t]
\centering
\vspace{-1.5em}
\includegraphics[width=0.8\linewidth]{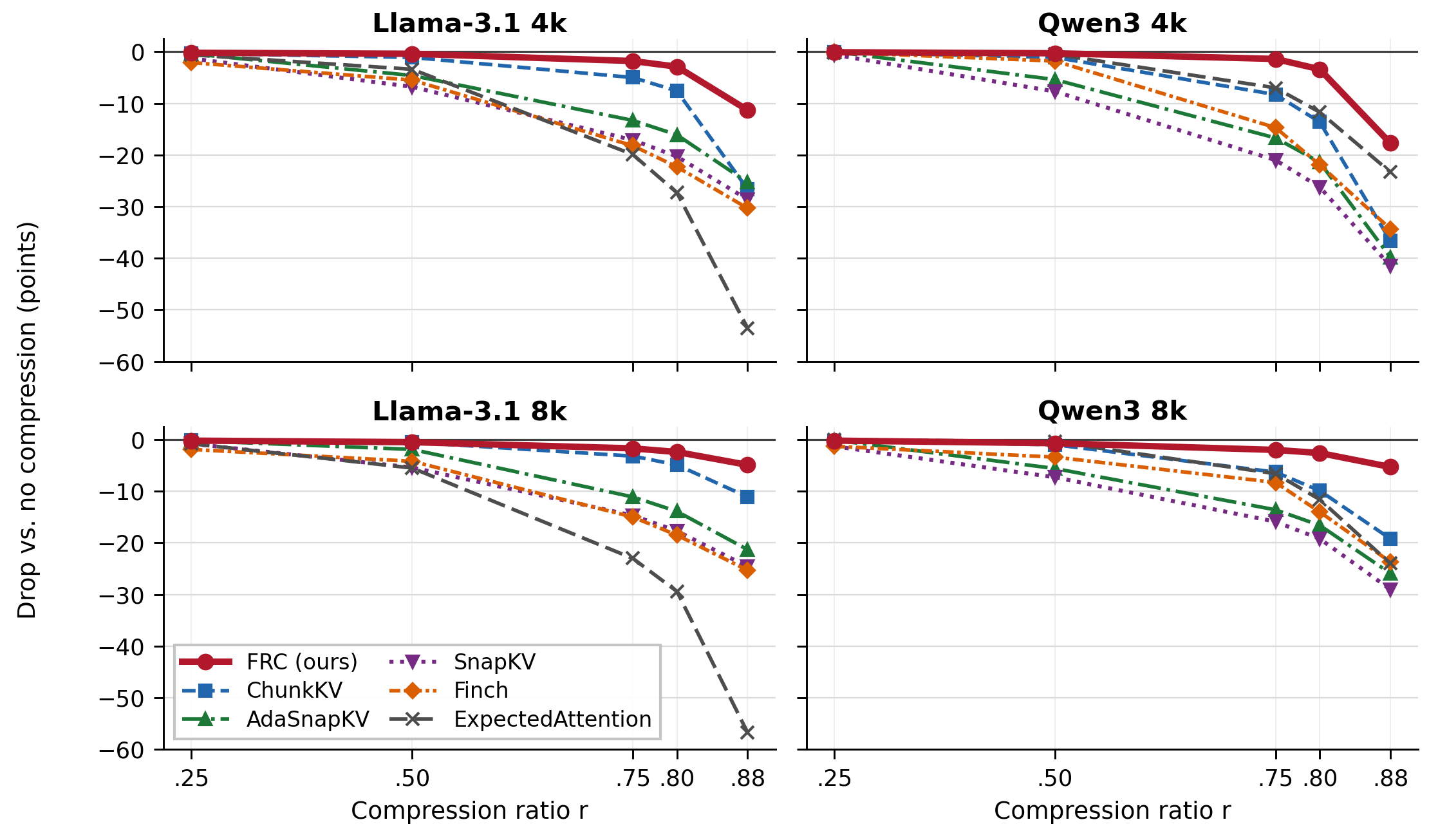}
\caption{RULER score drop relative to the no-compression upper bound across all 20 model/context/compression cells. Each panel corresponds to one model and context length, the $x$-axis is compression ratio, and lower curves indicate larger accuracy loss. The red FRC curve stays closest to zero throughout the aggressive-compression regime ($r \geq 0.75$), showing that FRC loses less accuracy than the evaluated single-pass baselines as the retained KV-cache budget shrinks.}

\label{fig:delta_baseline}
\vspace{-1.5em}
\end{figure}

The policy discovered by \Name is \textbf{FRC} (Feature-Rich Compression), a single-pass prefill-stage scorer that composes three MaxAbs-normalized attention signals: local attention received over a recent observation window, neighborhood attention density, and KV-head maximum salience. FRC uses fixed weights $(0.55, 0.30, 0.15)$ and chunk-level top-$k$ selection with chunk length 20. With the same weights and no per-model tuning, FRC ranks first among the evaluated single-pass KVPress baselines at every RULER~\citep{hsieh2024ruler} 4k/8k cell with $r \geq 0.75$ on both Llama-3.1-8B-Instruct~\citep{grattafiori2024llama} and Qwen3-8B~\citep{qwen3} (12 of 20 grid cells), gaining $+15.4$ points over ChunkKV at $88\%$ compression on Llama-4k and $+13.9$ points on Qwen-8k (Figure~\ref{fig:delta_baseline}).

A scorer-versus-structure decomposition isolates the mechanism. Replacing a V1-family scorer with FRC contributes $+67.2$ RULER points, while replacing chunk selection with tokenwise top-$k$ changes the result by only $\approx 0.1$. The load-bearing design choice is therefore the \emph{scoring family}, not the chunk-selection prior. The three FRC signals provide complementary views of token importance--temporal locality, spatial coherence, and head-specific salience--which explains why a fixed linear combination transfers across model families and context lengths better than any single attention prior. A constrained-archive analysis over $n{=}621$ linear-only programs from 13 evolution runs supports this interpretation: Spearman $\rho(\text{stage1}, \text{abs\_frc\_share}){=}0.513$, rising to $0.936$ above the noise floor, indicating that selection pressure within the editable surface moved toward FRC-aligned scoring.

Our contributions are three-fold:
(1) We formulate KV-cache eviction as editable-program search through \Name: a compact policy interface, cascade evaluation with strict output invariants, and a diagnostic loop for reformulating mis-scoped search surfaces.
(2) We discover FRC, a fixed-weight three-signal eviction scorer that transfers without per-model tuning and ranks first among evaluated single-pass KVPress baselines at every RULER 4k/8k cell with $r \geq 0.75$ across two GQA-equipped 8B models.
(3) We identify the main empirical mechanism: scoring formulation, not chunk structure, carries the gain; within FRC, \texttt{neighbor\_attn\_density} is the decisive signal-level addition.

We scope the claim to retrieval-style aggressive compression. FRC underperforms on long-document comprehension, where key-norm coverage is a stronger inductive bias, and we do not claim universal optimality or autonomous rediscovery from a structurally distant seed. We also do not evaluate two-pass methods such as KVzip~\citep{kim2025kvzip} or concurrent learned eviction methods such as KVP~\citep{moschella2026learning}, Locret~\citep{huang2024locret}, LookaheadKV~\citep{ahn2026lookaheadkv}, and Lookahead Q-Cache~\citep{wang2025lookahead}; these were not integrated into our harness and remain open comparisons.


\section{Background}
\label{sec:background}

\noindent\textbf{KV cache compression.}
LLM inference has a parallel prefill phase that builds the KV cache and an autoregressive decoding phase that repeatedly reads it. For an $L$-layer, $H$-head, $d$-dimensional model, cache memory scales linearly with context length as $\mathcal{O}(L H n d)$~\citep{pope2023efficiently,kwon2023vllm}. Grouped-Query Attention reduces the constant factor but does not remove the long-context memory bottleneck. This work focuses on \emph{prefill-stage eviction}: compressing the cache once, after the prompt forward pass and before generation begins.

\noindent\textbf{Single-pass prefill eviction.}
The prefill-stage methods we evaluate differ mainly in how they score token importance. SnapKV~\citep{li2024snapkv} pools attention over a recent observation window; ChunkKV~\citep{liu2025chunkkv} averages such scores over fixed chunks; AdaSnapKV~\citep{feng2024ada} reallocates budgets using entropy; Finch~\citep{corallo2024finch} adds query-aware fusion; and ExpectedAttention~\citep{devoto2025expected} uses key norms as a query-independent proxy. These methods are attractive because they are training-free and single-pass, but each commits to a hand-designed salience statistic. CacheCraft keeps the same deployment regime--single-pass, training-free prefill eviction--while searching over the scoring program itself.

\noindent\textbf{Program evolution for policy discovery.}
LLM-guided program evolution uses language-model edits over executable program text rather than hand-coded mutation operators, following systems such as FunSearch~\citep{romera2024mathematical}, AlphaEvolve~\citep{novikov2025alphaevolve}, and OpenEvolve~\citep{openevolve}. CacheCraft adapts this paradigm to KV-cache eviction by exposing a compact policy interface, enforcing output invariants during evaluation, and interpreting plateaus or reward-hacking failures as signals to reformulate the editable search surface.


\section{\Name Design}
\label{sec:method}

\Name turns prefill-stage KV-cache eviction into an executable program-search problem. A candidate policy is a small Python module that scores tokens from a fixed feature context and returns the indices to retain. The search loop has four steps: \textbf{formulate} the editable policy interface and reward; \textbf{seed} the search with a benchmark-credible prior; \textbf{evolve} candidate programs using an LLM-guided code-evolution engine; and \textbf{validate} surviving policies on held-out benchmark subsets before full evaluation. The central design principle is diagnostic: plateaus, invalid outputs, or reward-hacking failures are treated as evidence that the editable interface, seed family, or evaluator contract is mis-scoped, prompting reformulation rather than simply more iterations.

\begin{figure}[t]
\centering
\includegraphics[width=0.9\linewidth]{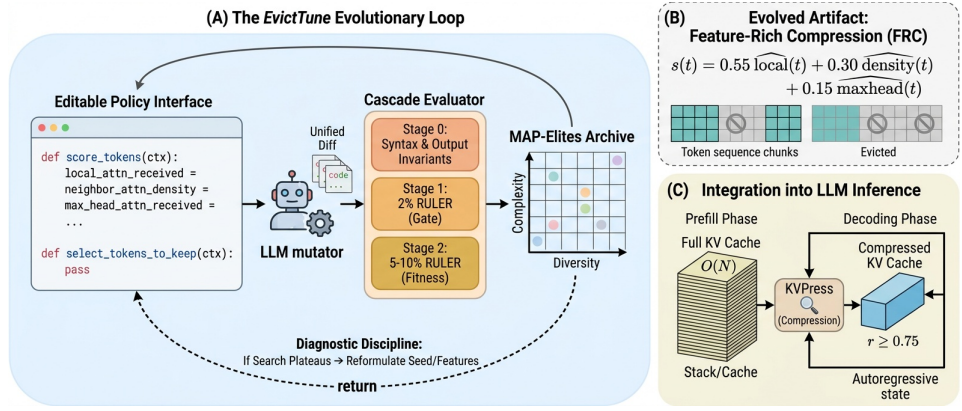}
\caption{\Name, FRC, and inference integration. \textbf{(A)~\Name loop:} an LLM mutator edits a compact KVPress policy; a cascade evaluator checks syntax, output contracts, and benchmark quality before admitting programs into an archive. Plateaus or reward-hacking failures prompt diagnostic reformulation of the seed, interface, or evaluator contract. \textbf{(B)~FRC artifact:} the discovered fixed-weight three-signal scorer with chunk-level top-$k$ ($L_c{=}20$). \textbf{(C)~Integration:} FRC runs once at the end of prefill; the compressed cache, with footprint $1{-}r$, feeds autoregressive decoding. \Name is the methodology, FRC the artifact, and OpenEvolve the underlying code-evolution engine.}
\label{fig:loop}
\end{figure}

\noindent\textbf{Editable policy interface.}
The mutable object is deliberately small: each candidate exposes \texttt{score\_tokens(ctx)} and \texttt{select\_tokens\_to\_keep(ctx)}. The context \texttt{ctx} is computed from a single attention pass and contains token-level salience, position, chunk, and metadata fields. This interface is expressive enough to represent literature-grounded KV eviction heuristics while constrained enough for automated edits to remain executable and auditable. Candidate policies are evaluated through the official KVPress runner; baselines use the same upstream evaluation path, while evolved policies are loaded through a thin prefill-press wrapper that calls the two candidate functions.

\noindent\textbf{Cascade evaluator and output contracts.}
Each candidate passes through a cascade: a Stage-0 syntax/import check, a Stage-1 small-fraction RULER gate, and a Stage-2 held-out evaluation whose score enters the archive. The exact Stage-1/Stage-2 fractions are run-specific and are tabulated in Appendix~\ref{app:configs}; the FRC main run used $0.02/0.05$, while scaled redux runs used larger Stage-2 subsets. The archive fitness is $0.25\,S_1 + 0.75\,S_2$, where $S_1$ and $S_2$ are Stage-1 and Stage-2 aggregate scores.

The evaluator enforces the output contract before assigning quality. For tokenwise policies, \texttt{select\_tokens\_to\_keep} must return exactly \texttt{cache\_budget} unique indices, all in $[0,L)$. For chunk-structured policies such as ChunkKV and FRC, the retained length is rounded to whole chunks; the required contract is that returned indices are unique, in range, non-empty, and have a common retained length across batch items, with the retained count determined by the chunk selector rather than exact equality to $\lfloor L(1-r)\rfloor$. Violations are classified as \emph{invalid output}, not low quality, so no-op policies that keep the full cache cannot appear high-scoring.

This contract was strengthened after a mutation in a scaled redux run dropped the \texttt{[-k:]} slice from a selection wrapper, causing the candidate to return all indices and receive an artificially high score (Appendix~\ref{app:redux}). In \Name, such reward hacking is treated as a missing evaluator invariant: the remedy is to harden the contract, not to discard the search space.

\noindent\textbf{LLM-guided evolution and diagnostic reformulation.}
We instantiate the search using OpenEvolve~\citep{openevolve}, which maintains an archive of candidate programs and proposes mutations as code edits. The main paper uses OpenEvolve as the engine; \Name's contribution is the KV-specific formulation: the editable policy surface, the cascade evaluator, the output contracts, and the diagnostic reformulation loop. Implementation parameters--archive dimensions, parent sampling, island topology, mutator checkpoint, and per-run budgets--are reported in Appendix~\ref{app:configs}.

The diagnostic loop is what connects search behavior back to methodology. A plateau suggests that the exposed features or seed family are too weak; a cross-ratio failure suggests the search operating point is misaligned with the target regime; a reward-hacking incident suggests that the evaluator contract is incomplete. Section~\ref{sec:formulation} shows how this loop led to the FRC policy, while Appendix~\ref{app:algorithms} gives executable-style pseudocode for the evolutionary loop, cascade evaluator, feature construction, chunk selection, and GPU-native FRC \texttt{compress()} implementation.

\noindent\textbf{Feature context.}
The policy context exposes 17 token-level fields and 5 metadata fields covering global attention statistics, local and neighborhood density, per-head salience, key-space features, geometry/position features, chunk summaries, boolean masks, and scalar budget metadata. Continuous attention-, key-, and chunk-mean features are MaxAbs-normalized; position, rank, mask, and scalar fields remain in their native bounded scales. Closed-form definitions are in Appendix~\ref{app:features}.

============================================================================
\section{The Discovered Policy: FRC}
\label{sec:formulation}

\textbf{Policy form.}
\Name discovers \textbf{FRC} (Feature-Rich Compression), a single-pass prefill-stage eviction policy that combines a fixed three-signal scorer with chunk-level top-$k$ selection. Given token scores $s(t)$, FRC averages scores within chunks of length $L_c{=}20$ and retains the top-scoring chunks. This chunk structure is adopted from ChunkKV, but used as a scorer-modular selection prior: Section~\ref{sec:results_ablation} ablates chunk selection independently from the scoring family.

The discovered scorer composes three attention-derived signals: \texttt{local\_attn\_received}, a recent observation-window score; \texttt{neighbor\_attn\_density}, a width-adaptive moving average of mean received attention; and \texttt{max\_head\_attn\_received}, a KV-head salience maximum. These signals represent temporal locality, spatial coherence, and head-specific salience, respectively. The same fixed weights and width parameters are used in every evaluation cell, with no per-model tuning.

\textbf{Scoring rule.}
For GQA models, FRC computes attention-derived features from the KV-head attention proxy in Listing~\ref{alg:proxy}. If $G=H/H_{\mathrm{kv}}$ query heads share each KV head, define
\[
\widetilde A_{g,q,t}=\frac{1}{G}\sum_{j=1}^{G}A_{g,j,q,t},
\]
where $g$ indexes KV heads and $j$ indexes query heads within the group. The final scorer is
\begin{equation}
 s(t)=0.55\;\widehat{\mathrm{local}}(t)+0.30\;\widehat{\mathrm{density}}(t)+0.15\;\widehat{\mathrm{maxhead}}(t),
\label{eq:score}
\end{equation}
where $\mathrm{local}(t)=\tfrac{1}{H_{\mathrm{kv}}W}\sum_g\sum_{q=Q-W}^{Q-1}\widetilde A_{g,q,t}$ with $W{=}32$, $\mathrm{density}(t)$ is a moving average of $\bar a(t)=\tfrac{1}{H_{\mathrm{kv}}Q}\sum_{g,q}\widetilde A_{g,q,t}$ using $W_d=\max(3,\min(33,2\lfloor L/64\rfloor+1))$, and $\mathrm{maxhead}(t)=\max_g\tfrac{1}{Q}\sum_q\widetilde A_{g,q,t}$. Thus max-head salience is the maximum over KV-group-averaged heads, not raw individual query heads. Hats denote per-token MaxAbs normalization.

\textbf{How the diagnostic loop arrived at FRC.}
The V1 through V5 trajectory (Table~\ref{tab:v1v5}) illustrates how \Name reformulates the search surface when evolution plateaus. V1 used a tokenwise five-signal scorer over global attention, tail attention, head consistency, key norms, and position; at mild compression ($r{=}0.12$), it reached 76.9. Re-seeding V2 and V3 improved this to 83.0 and then 83.6, a plateau that implicated the seed family rather than the iteration budget. V4 added query awareness and reached 85.5 at the same mild ratio, but a cross-ratio sweep exposed the failure mode: V4 fell roughly 20 points behind baselines at $r{=}0.25$ and 40 points behind at $r{=}0.83$. We therefore changed both the operating point and the feature surface, evolving V5 at $r{=}0.83$ with a literature-guided feature audit over 17 candidate signals (Appendix~\ref{app:features}) and the chunk-level selection prior above. None of the five V1 signals survives into the V5 policy family; the loop substantially reformulated the editable surface rather than merely tuning weights.

\begin{table}[t]
\centering
\caption{V1 through V5 trajectory at a glance. Scores are RULER 4k aggregates; signal sets are the editable feature columns exposed at each stage. Diagnostic interventions (D1, D2) mark points where the loop reformulated rather than iterated. The V5 seed scored $\approx 91.2$ before evolution (weights $(0.60,0.25,0.15)$); the V5 run refined the weights to $(0.55,0.30,0.15)$, adding $\approx 0.2$.}
\label{tab:v1v5}
\small
\begin{tabular}{lllll}
\toprule
Stage & Operating point & Editable surface & RULER 4k & Diagnostic outcome \\
\midrule
V1 & $r{=}0.12$, tokenwise   & 5 V1-family signals       & 76.9          & --- \\
V2 & $r{=}0.12$, tokenwise   & V1 family, re-seeded      & 83.0          & --- \\
V3 & $r{=}0.12$, tokenwise   & V1 family, re-seeded      & 83.6 (plateau) & \textbf{D1: interface mis-scoped} \\
V4 & $r{=}0.12$, query-aware & V1 family + query         & 85.5          & \textbf{D2: cross-ratio failure} \\
V5 & $r{=}0.83$, chunk-level & FRC 3-signal family       & $\approx 91.2 \to 91.4$ & loop converged \\
\bottomrule
\end{tabular}
\end{table}

The V5 seed already scored $\approx 91.2$ on RULER 4k at $r{=}0.83$ with weights $(0.60,0.25,0.15)$; evolution refined the weights to $(0.55,0.30,0.15)$ and added $\approx 0.2$. This small final improvement suggests that, by V5, the diagnostic loop had found the right interface and seed family, leaving weight refinement as the remaining work. Appendix~\ref{app:evobounds} bounds this claim.

\section{Experimental Setup}
\label{sec:setup}

We evaluate two GQA-equipped 8B-parameter instruction-tuned models, Llama-3.1-8B-Instruct~\citep{grattafiori2024llama} and Qwen3-8B~\citep{qwen3}, on RULER~\citep{hsieh2024ruler} at $4{,}096$- and $8{,}192$-token contexts. We report
the RULER aggregate used by our evaluation harness: the unweighted arithmetic mean over the 13 individual RULER subtasks. For compact reporting, Appendix~\ref{app:full_tables} groups these subtasks into eight task-family columns (single-needle, multi-needle, multi-key, multi-value, variable tracking, common-word extraction, frequent-word extraction, and QA), but the reported AVG is computed over the 13 raw subtasks rather than equally over the eight grouped families. NIAH at $16{,}384$ tokens is in Appendix~\ref{app:additional}. We sweep compression ratios $r \in \{0.25, 0.50, 0.75, 0.80, 0.88\}$, where $r$ is the fraction of the KV cache discarded and the retained budget is $(1{-}r) L$. Prefill-stage baselines come from KVPress and are restricted to single-pass methods: ChunkKV ($C{=}20$), SnapKV, AdaSnapKV, Finch (query-aware), and ExpectedAttention (key-norm-based)~\citep{kvpress}, plus a no-compression upper bound.

Decoding is greedy and deterministic. Evaluation runs used NVIDIA H100 (80\,GB) and H200 (141\,GB) GPUs depending on availability. Appendix~\ref{app:latency} reports a separate H100-only latency microbenchmark for the measured FRC, ChunkKV, and SnapKV implementations.
The discovery loop uses OpenEvolve v0.5.1 behind a multi-stage cascade: Stage-1 candidates above a quality threshold are promoted to Stage-2, and only Stage-2 survivors enter the archive. Appendix~\ref{app:configs} tabulates per-iteration hyperparameters, including the mutator checkpoint, cascade subset fractions, gating thresholds, and per-run iteration budgets.

\textbf{Scope.} We restrict the evaluated baselines to single-pass prefill-stage methods. Two-pass methods (notably KVzip~\citep{kim2025kvzip}) were not run end-to-end: their KVPress integration was not active in our local environment, and KVzip requires a second prefill pass that our harness does not issue. We therefore report them as open comparisons. Concurrent prefill-stage baselines cited in Section~\ref{sec:related} (KVCompose, LAVa, Compactor, CurDKV, EvolKV) are acknowledged but not integrated into our harness. LongBench-v2~\citep{bai2025longbench} is reported for Llama-3.1-8B-Instruct only; Qwen3-8B is excluded from the LongBench-v2 headline because its no-compression baseline was near the random-guess floor, leaving no meaningful compression signal to compare. 


\section{Main Experimental Results}
\label{sec:results}

\begin{wrapfigure}{r}{0.45\textwidth}
\centering
\vspace{-1.5em}
\includegraphics[width=\linewidth]{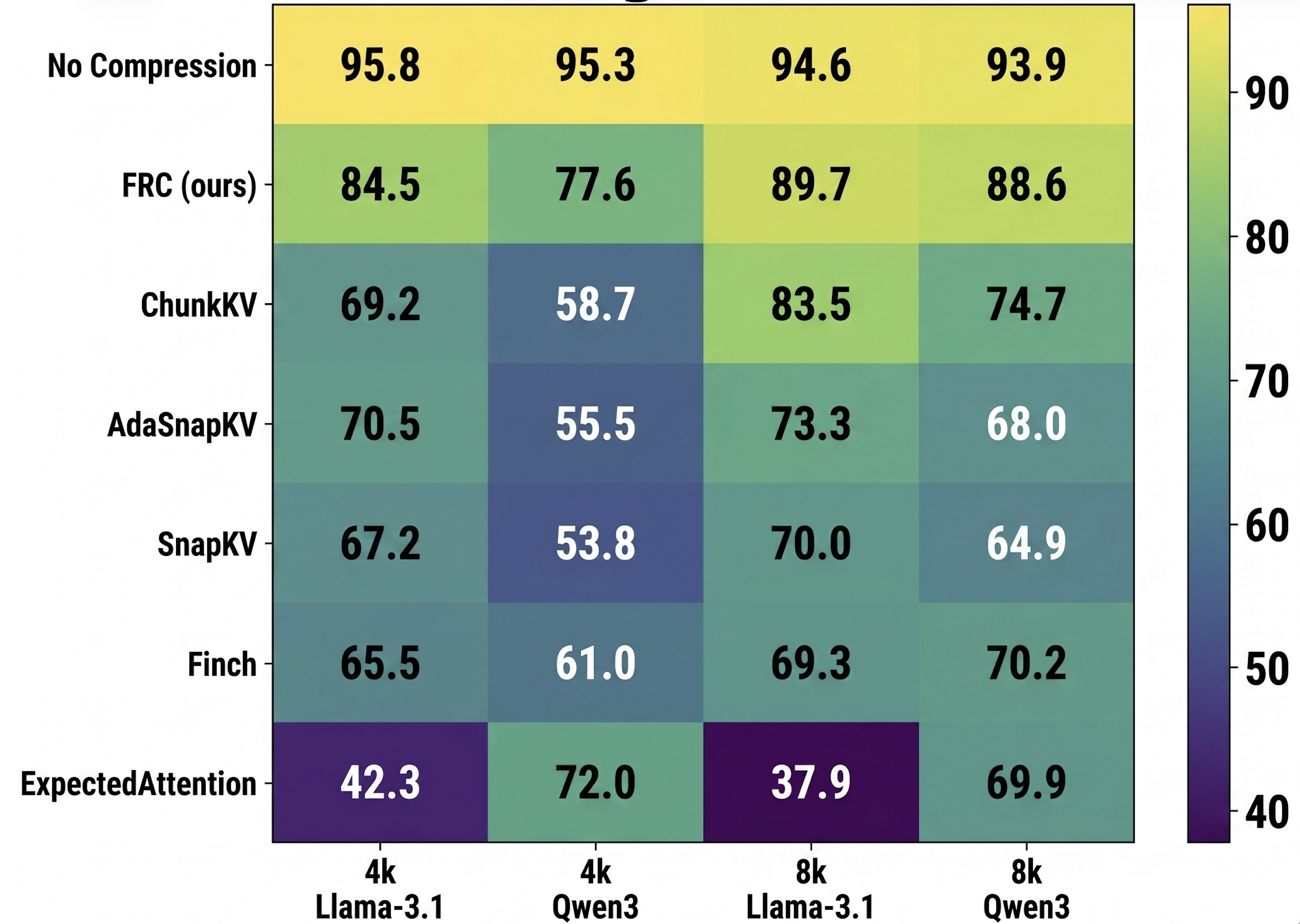}
\caption{RULER aggregate accuracy at $r{=}0.88$. FRC ranks second only to no compression in every model/context cell using identical fixed weights.}
\vspace{-2.0em}
\label{fig:heatmap_r088}
\end{wrapfigure}

We evaluate whether FRC satisfies the design goal from Section~\ref{sec:formulation}: a fixed, training-free eviction scorer that transfers across model families, context lengths, and aggressive compression ratios without per-model retuning. The main result is that FRC is consistently strongest among the evaluated single-pass KVPress baselines in the high-compression regime, and the ablations show that this gain comes from the scoring family rather than the chunk-selection structure. Non-RULER audits are treated as limitations rather than headline evidence and are discussed in Section~\ref{sec:limits}.

Figure~\ref{fig:heatmap_r088} shows the headline operating point. At $r{=}0.88$, the same fixed weights $(0.55,0.30,0.15)$ place FRC second only to no compression in every (model, context) cell. Across the full $2 \times 2 \times 5$ RULER sweep, FRC ranks first among compression methods at all 12 cells with $r \geq 0.75$; the eight lower-compression cells with $r \leq 0.50$ are tightly clustered and should not be interpreted as robust separations.

\subsection{RULER performance under compression}
\label{sec:results_ruler}

Figure~\ref{fig:delta_baseline} reports the full sweep as the drop from the no-compression upper bound. At 4k context and $r{=}0.88$, FRC reaches 84.6 on Llama-3.1-8B-Instruct versus 69.2 for the strongest baseline (ChunkKV), a $+15.4$ point gap. On Qwen3-8B at the same context and compression ratio, FRC reaches 77.6 versus 72.0 for the strongest baseline (ExpectedAttention), a $+5.6$ point gap. The strongest non-FRC baseline therefore reorders across models, while FRC remains rank-1 among compressed methods.

At 8k context, the same pattern holds with the same FRC weights. On Qwen3-8B at $r{=}0.88$, FRC reaches 88.6 versus 74.7 for ChunkKV, the largest single-cell gap in the paper ($+13.9$ points). On Llama-3.1-8B-Instruct, FRC reaches 89.7 versus 83.5 for ChunkKV, a $+6.2$ point gap. Gaps narrow at $r{=}0.75$ for longer contexts (2048 retained tokens at 8k versus 1024 at 4k), but FRC remains first among compressed methods at every high-compression cell.

\begin{figure*}[t]
  \centering
  \begin{subfigure}[t]{0.45\linewidth}
    \centering
    \includegraphics[width=\linewidth]{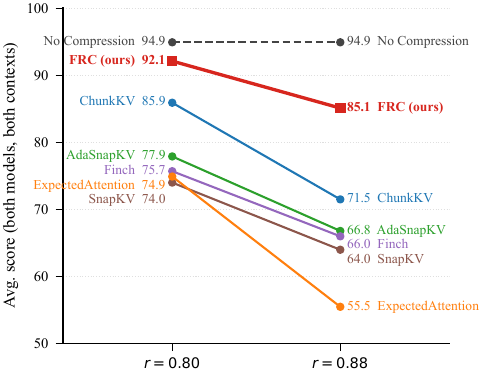}
    \caption{High-compression slopegraph. Average score across both
      models and both context lengths at $r{=}0.80$ and $r{=}0.88$.
      FRC retains $85.1$ at $r{=}0.88$, a $13.6$-point gap over the
      next baseline (ChunkKV, $71.5$) and a $29.6$-point gap over
      ExpectedAttention.}
    \label{fig:slopegraph}
  \end{subfigure}\hfill
  \begin{subfigure}[t]{0.45\linewidth}
    \centering
    \includegraphics[width=\linewidth]{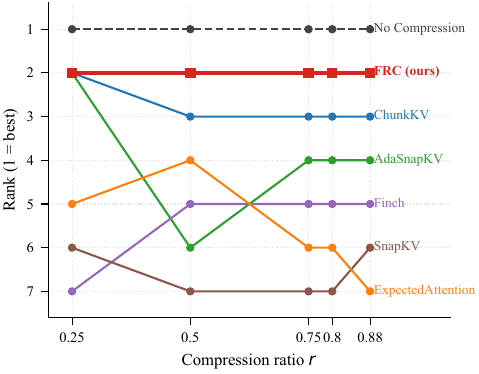}
    \caption{Rank trajectory across compression ratios
      $r\in\{0.25,0.50,0.75,0.80,0.88\}$ ($1$ = best).
      FRC is the only KV-compression method that holds rank~2 at every
      ratio; baselines reorder under pressure.}
    \label{fig:rank}
  \end{subfigure}
  \caption{KV-cache compression behavior under increasing pressure.
    \textbf{(a)} Absolute scores at the two most aggressive ratios.
    \textbf{(b)} Method ranks across the full sweep.
    FRC dominates baselines at every ratio and degrades the slowest
    in the high-compression regime.}
  \label{fig:compression-overview}
\end{figure*}

Figure~\ref{fig:slopegraph} summarizes compression sensitivity. Between $r{=}0.80$ and $r{=}0.88$, FRC drops by $\approx 7$ points averaged over both models and contexts, while ChunkKV drops by $\approx 14$ and ExpectedAttention by $\approx 19$. This indicates that FRC's advantage is not merely a single-cell artifact; it degrades more slowly as the retained cache budget becomes small.

\subsection{What drives the gain: scorer versus structure}
\label{sec:results_ablation}

The central ablation asks whether FRC's gains come from ChunkKV-style chunk selection or from the discovered scoring family. We hold the cascade and budget fixed and vary the scoring family and selection structure independently. The test uses $r{=}0.88$ on Llama-3.1-8B-Instruct, where only 12\% of tokens are retained and scoring quality is most exposed.

\begin{table}[t]
\centering
\caption{Scorer-versus-structure decomposition at $r{=}0.88$ on Llama-3.1-8B-Instruct, RULER aggregate. Holding the cascade and selection budget fixed, we vary the scoring family (rows a$\to$b and a$\to$c) and the selection structure (rows b$\to$c) independently.}
\label{tab:ablation}
\small
\begin{tabular}{llrr}
\toprule
& Configuration & RULER agg. & $\Delta$ vs.\ (a) \\
\midrule
(a) & V1-family scorer + tokenwise top-$k$ & 17.4 & -- \\
(b) & FRC scorer + chunk structure (V5 evolved policy) & 84.6 & $+67.2$ \\
(c) & FRC scorer + tokenwise top-$k$ & 84.7 & $+67.3$ \\
\bottomrule

\end{tabular}
\end{table}

Table~\ref{tab:ablation} isolates the mechanism. Replacing chunk selection with tokenwise top-$k$ changes the score by only $\approx 0.1$ (row b$\to$c), while replacing the V1-family scorer with FRC changes the score by $+67$ points (row a$\to$b/c). The chunk structure is therefore not the load-bearing component; the scoring family is. This rules out the interpretation that FRC is merely ChunkKV with small tweaks.

A second ablation holds chunk structure fixed and varies the scorer \emph{within} the FRC family (Table~\ref{tab:within_frc}). Removing \texttt{neighbor\_attn\_density} costs $-3.20$ points, removing \texttt{max\_head\_attn\_received} costs only $-0.22$, and the recency-only single-signal scorer (87.49) already beats official ChunkKV (85.10). The decisive within-FRC addition is \texttt{neighbor\_attn\_density}, which aligns local attention with chunk granularity by rewarding coherent salient regions; \texttt{max\_head\_attn\_received} provides a smaller refinement.

\begin{table}[t]
\centering
\caption{Within-FRC signal-level ablation at $r{=}0.83$ on Llama-3.1-8B-Instruct, full-scale RULER 4k aggregate. Chunk structure and chunk length $L_c{=}20$ held fixed; only the scoring formula varies.}
\label{tab:within_frc}
\small
\begin{tabular}{l l r r}
\toprule
Variant & Scorer formula & RULER agg.\ & $\Delta$ vs.\ FRC \\
\midrule
FRC (full) & $0.55\,\widehat{\text{local}} + 0.30\,\widehat{\text{neighbor}} + 0.15\,\widehat{\text{maxhead}}$ & 91.23 & -- \\
$-$ \texttt{max\_head} & $0.65\,\widehat{\text{local}} + 0.35\,\widehat{\text{neighbor}}$ & 91.01 & $-0.22$ \\
$-$ \texttt{neighbor} & $0.79\,\widehat{\text{local}} + 0.21\,\widehat{\text{maxhead}}$ & 88.03 & $-3.20$ \\
local-only & $1.00\,\widehat{\text{local}}$ & 87.49 & $-3.74$ \\
\bottomrule
\end{tabular}
\end{table}

\section{Analysis and discussion}\label{sec:analysis}

\textbf{Why three signals compose well.}
The $+67.2$ versus $\approx 0.1$ decomposition in \S\ref{sec:results_ablation} attributes the gain to the scoring family, not the selection structure. Within that family, Table~\ref{tab:within_frc} shows that each signal captures a distinct mode of token importance. \texttt{local\_attn\_received} provides recent-window salience; \texttt{neighbor\_attn\_density} smooths noisy token scores into a chunk-aligned salience map and is the largest within-family contributor ($-3.20$ when removed); and \texttt{max\_head\_attn\_received} preserves sparse head-specific evidence that averaging can erase ($-0.22$ when removed). These axes--temporal locality, spatial coherence, and head heterogeneity--explain why a fixed three-way scorer transfers across both models better than any single salience prior.

\textbf{What the search dynamics imply.}
The diagnostic loop matters because the final FRC run was not a brute-force rediscovery from an arbitrary starting point. As Section~\ref{sec:formulation} shows, the loop first had to reformulate the editable interface and operating point before evolution could refine the final scorer; Appendix~\ref{app:evobounds} bounds what evolution did within V5. Four diversity-controlled OpenEvolve runs from a structurally distant V1 five-signal seed (varying \texttt{num\_islands}, mutator temperature, and mutator LLM, $500$ iterations each at $r{=}0.83$) all stayed within the V1 family rather than discovering FRC's family. This suggests that local program mutation alone does not reliably bridge large representation gaps; the editable surface must expose the right abstractions.

At the same time, selection pressure within a well-scoped surface is meaningful. A constrained-archive analysis over $n{=}621$ linear-only programs from 13 runs finds Spearman $\rho(\text{stage1}, \text{abs\_frc\_share}){=}0.513$, rising to $0.936$ above the noise floor, with a positive per-generation FRC-share trajectory in 11 of 11 runs. Thus \Name's contribution is not autonomous discovery from any seed, but a disciplined loop: diagnose plateaus or reward-hacking failures, reformulate the interface or evaluator contract, and then let evolution search within that scoped surface. Per-run details, the B-Redux reward-hacking incident, and archive analysis are in Appendix~\ref{app:redux}.

\section{Limitations}
\label{sec:limits}

\textbf{Baseline and coverage frontier.}
Our comparisons cover the single-pass prefill-stage KVPress baselines operational in our local cascade. Two-pass methods (KVzip~\citep{kim2025kvzip}), concurrent prefill-stage methods (KVCompose, LAVa, Compactor, CurDKV, EvolKV, PagedEviction~\citep{chitty2026pagedeviction}), and concurrent learning-based methods (KVP~\citep{moschella2026learning}, Locret~\citep{huang2024locret}, LookaheadKV~\citep{ahn2026lookaheadkv}, Lookahead Q-Cache~\citep{wang2025lookahead}) are cited but not integrated. We therefore do not claim a global accuracy--latency Pareto frontier against two-pass reconstruction methods or learned eviction policies. Our latency appendix isolates implementation overhead only relative to the measured single-pass baselines: FRC-GPU is faster than ChunkKV in our harness but slower than the single-signal SnapKV reference, as expected from computing three salience signals rather than one.

\textbf{Evaluation scope and transfer.}
We evaluate two GQA-equipped 8B instruction models, RULER 4k/8k, NIAH 16k, and exploratory InfiniteBench and LongBench-v2 audits (Appendices~\ref{app:additional} and~\ref{app:longbench}); RULER 16k, 70B-class models, and non-GQA architectures remain open. FRC is not universally dominant. On LongBench-v2~\citep{bai2025longbench} at $r{=}0.83$ on Llama-3.1-8B-Instruct, ExpectedAttention is stronger ($0.264$ vs.\ $0.187$). The distinction is regime-based: FRC is retrieval-oriented, while long-document comprehension benefits from key-norm coverage. The framework's prescription is benchmark-specific re-evolution from the same seed contract rather than universal weights.

FRC also uses fixed absolute smoothing windows ($W{=}32$, $W_d{\leq}33$). These constants are held fixed across the evaluated context range, but we do not claim they are optimal for ultra-long contexts such as 128k tokens, where their relative span is much smaller; future runs should expose window size or multi-scale density features to the evolutionary search rather than treating these constants as universal.

\noindent\textbf{Methodological controls.}
A static random-feature linear baseline that would separate ``three normalized attention signals at fixed weights'' from ``the FRC three signals specifically'' is future work; the closest in-body proxy is row~(a) of Table~\ref{tab:ablation}, where a non-FRC V1-family scorer scores 17.4 versus FRC's 84.6. We used \texttt{qwen3.5:35b-a3b-q4\_K\_M}; AlphaEvolve-class frontier mutators may traverse wider regions within shorter budgets.

\noindent\textbf{Broader impacts.}
FRC is an inference-efficiency method: its positive impact is reduced KV-cache memory pressure and lower serving cost for long-context models. As with other efficiency improvements, reducing inference cost can also lower the cost of harmful uses of existing language models. The method does not introduce new model capabilities, training data, or a user-facing system; deployment risks remain governed by the underlying model and application context.

\section{Related Works}
\label{sec:related}

\noindent\textbf{KV-cache compression methods.}
Decoding-phase methods such as StreamingLLM~\citep{xiao2023efficient}, H2O~\citep{zhang2023h2o}, and PyramidKV~\citep{cai2024pyramidkv} target a different operating point from our prefill-stage setting and are complementary. Within prefill-stage eviction, single-pass methods such as SnapKV~\citep{li2024snapkv}, ChunkKV~\citep{liu2025chunkkv}, AdaSnapKV~\citep{feng2024ada}, Finch~\citep{corallo2024finch}, and ExpectedAttention~\citep{devoto2025expected} provide the baselines in our harness. Concurrent prefill-stage work broadens this design space through composite-token selection, residual-stream budget allocation, leverage-style scoring, or systems-level paging, including KVCompose, LAVa, Compactor, CurDKV, and PagedEviction~\citep{chitty2026pagedeviction}. Two-pass reconstruction methods such as KVzip~\citep{kim2025kvzip} fall outside our single-pass harness and remain open comparisons.

\noindent\textbf{Learned eviction policies.}
KVP~\citep{moschella2026learning}, Locret~\citep{huang2024locret}, LookaheadKV~\citep{ahn2026lookaheadkv}, and Lookahead Q-Cache~\citep{wang2025lookahead} learn model-specific retention policies through reinforcement learning, supervised fine-tuning, or lookahead-query prediction. Their design point is complementary to FRC: they may exploit model-specific training signals, whereas FRC is training-free, fixed-weight, and used without per-model adaptation.

\noindent\textbf{Evolutionary search for KV policies.}
The closest concurrent work is EvolKV~\citep{yu2025evolkv}, which also applies evolutionary search to KV-cache compression but searches a different object. EvolKV evolves a real-valued vector of per-layer cache budgets on top of an existing eviction rule. \Name instead evolves the eviction rule itself as editable program text over a structured feature interface. Thus EvolKV searches allocation, while \Name searches scoring logic; the two approaches are orthogonal and could be combined.
Overall, FRC occupies the training-free, single-pass prefill regime, while \Name contributes the search formulation that discovers such policies rather than hand-designing them.



\section{Conclusion}
\label{sec:conclusion}

We presented \Name, a program-evolution methodology for discovering prefill-stage KV-cache eviction policies, and FRC, the fixed-weight three-signal policy discovered by this process. Across two 8B GQA models and RULER 4k/8k contexts, FRC ranks first among the evaluated single-pass KVPress baselines at every high-compression cell with $r \geq 0.75$. The main empirical lesson is that scoring formulation, not chunk structure, carries the gain: replacing the scorer contributes $+67.2$ RULER points, while changing the selection structure contributes only $\approx 0.1$. The methodological lesson is that LLM-guided program evolution is most useful when paired with a compact editable interface, strict evaluator invariants, and diagnostic reformulation of failed searches. The supplemental release contains configs (App.~\ref{app:configs}), the editable feature module (App.~\ref{app:features}), per-task tables (App.~\ref{app:full_tables}), and a seed/config/code-hash-keyed reproduction harness; anonymized code is included in the supplemental materials and the public repository will be released upon deanonymization.

\bibliographystyle{plainnat}
\bibliography{references}

\appendix

\section{Full RULER tables}
\label{app:full_tables}

This appendix tabulates the headline numbers from Section~\ref{sec:results}.

The per-task tables group RULER scores into eight task-family columns for readability: single-needle (S\textsubscript{1-3}), multi-needle (multi-query), multi-key (MK\textsubscript{1-3}), multi-value, variable tracking (VT), common-word extraction (CWE), frequent-word extraction (FWE), and QA (QA\textsubscript{1-2}). The AVG column, however, is the unweighted arithmetic mean over the 13 individual RULER subtasks emitted by the evaluation harness, not an equal-weighted mean over the eight grouped columns. Methods compared are FRC (ours), ChunkKV, SnapKV, AdaSnapKV, Finch, ExpectedAttention~\citep{kvpress,li2024snapkv,liu2025chunkkv,feng2024ada,corallo2024finch,devoto2025expected}, and the no-compression upper bound. Compression ratios are $r \in \{0.25, 0.50, 0.75, 0.80, 0.88\}$. All numbers come from official KVPress~\citep{kvpress} runs on RULER~\citep{hsieh2024ruler} with greedy decoding, \texttt{seed=42}, and \texttt{fraction=1.0}. Tables A.1 and A.2 give per-task scores at 4k context for Llama-3.1-8B-Instruct~\citep{grattafiori2024llama} and Qwen3-8B~\citep{qwen3}, respectively; Tables A.3 and A.4 give the per-method aggregate at 8k context (the per-task 8k breakdown is in the auxiliary release artifact).

\subsection{Absolute-score line plots and high-compression slopegraph}
\label{app:abs_lineplots}

\begin{figure}[ht]
\centering
\includegraphics[width=0.95\linewidth]{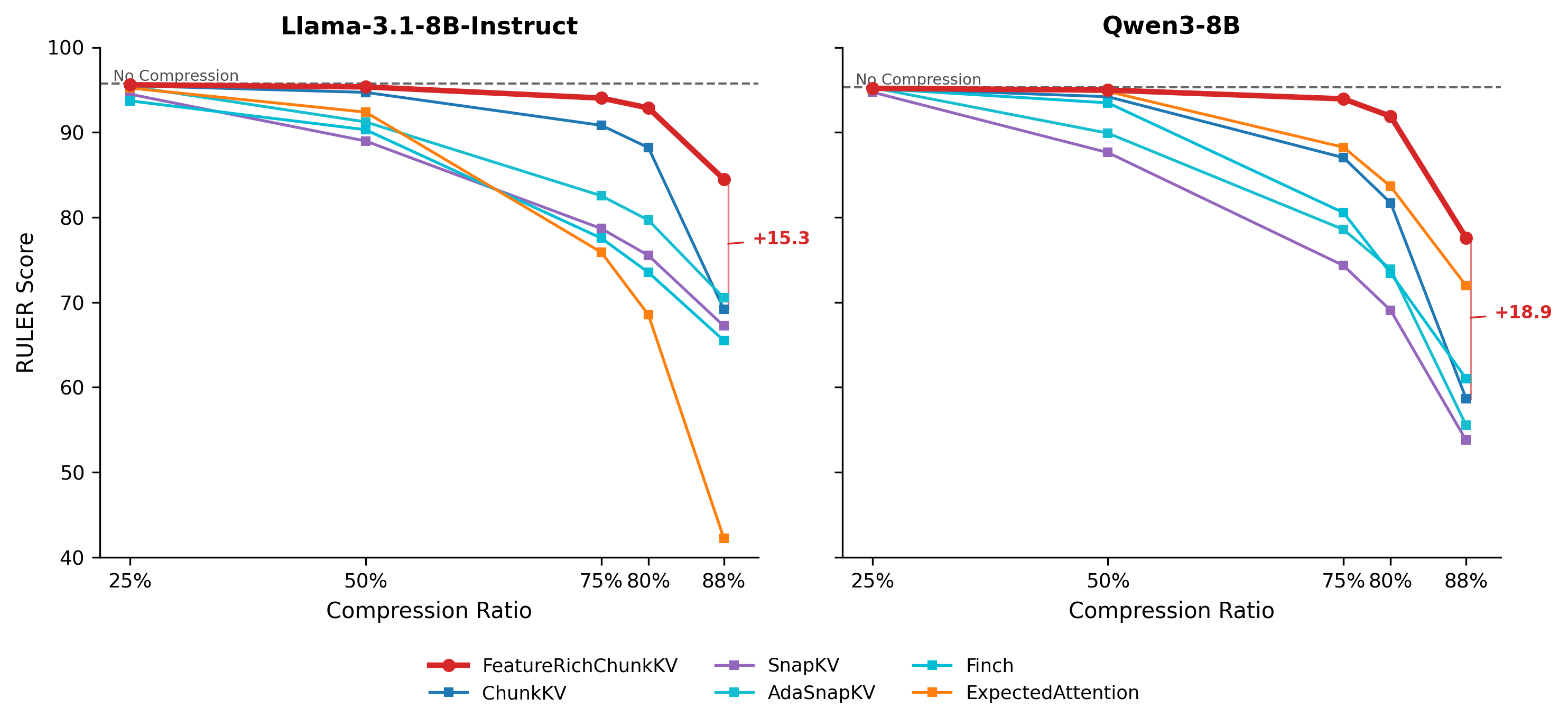}
\caption{RULER 4k accuracy versus compression ratio for the two evaluated models. FRC ranks first at every cell with $r \geq 0.75$ on both panels. Baseline rankings reorder substantially between Llama and Qwen at $r{=}0.88$: ChunkKV is the strongest non-FRC method on Llama, whereas ExpectedAttention is strongest on Qwen. This is the absolute-score view of the left half of main-text Figure~\ref{fig:delta_baseline}.}
\label{fig:app_ruler4k}
\end{figure}

\begin{figure}[ht]
\centering
\includegraphics[width=0.95\linewidth]{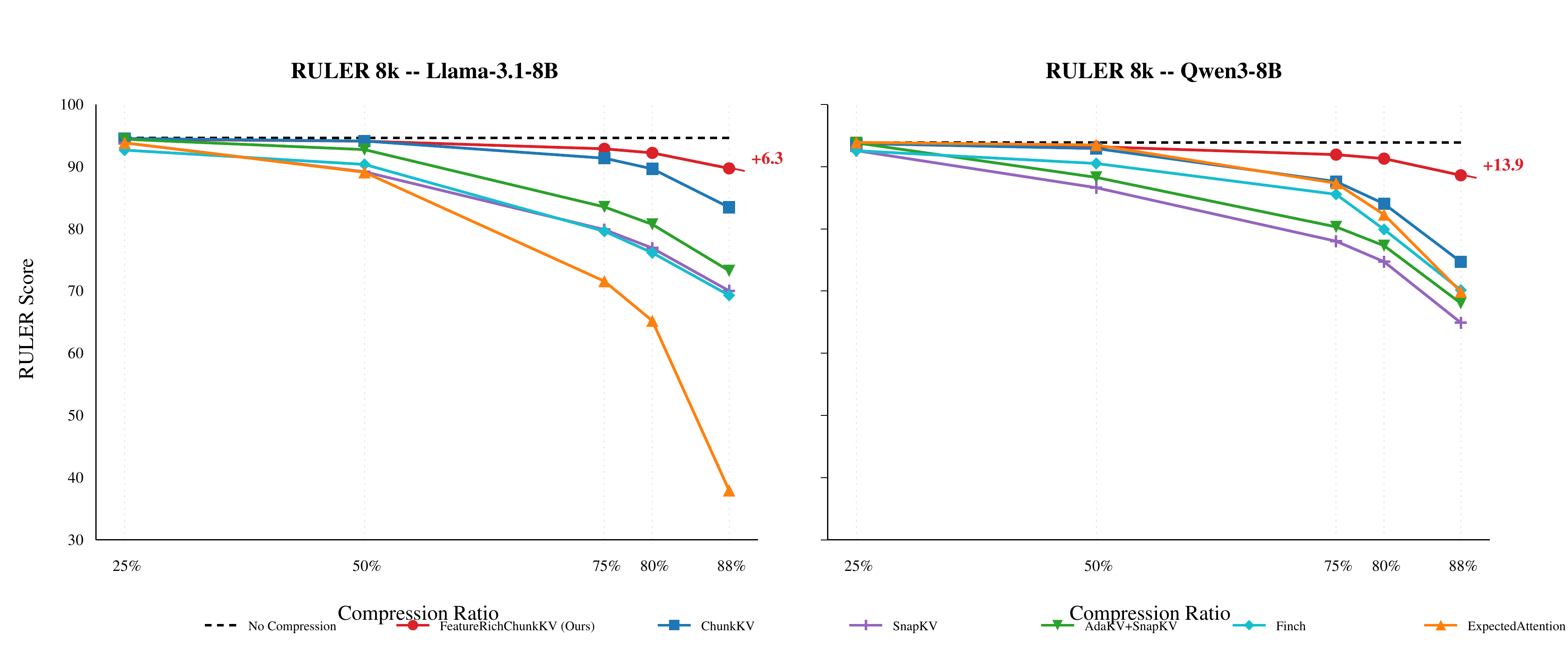}
\caption{RULER 8k accuracy versus compression ratio. FRC remains first at every cell with $r \geq 0.75$ on both models using the same fixed weights $(0.55, 0.30, 0.15)$ as at 4k. The Qwen 8k gap at $r{=}0.88$ widens to $+13.9$ over the strongest baseline (ChunkKV). This is the absolute-score view of the right half of main-text Figure~\ref{fig:delta_baseline}.}
\label{fig:app_ruler8k}
\end{figure}

\begin{figure}[ht]
\centering
\includegraphics[width=0.95\linewidth]{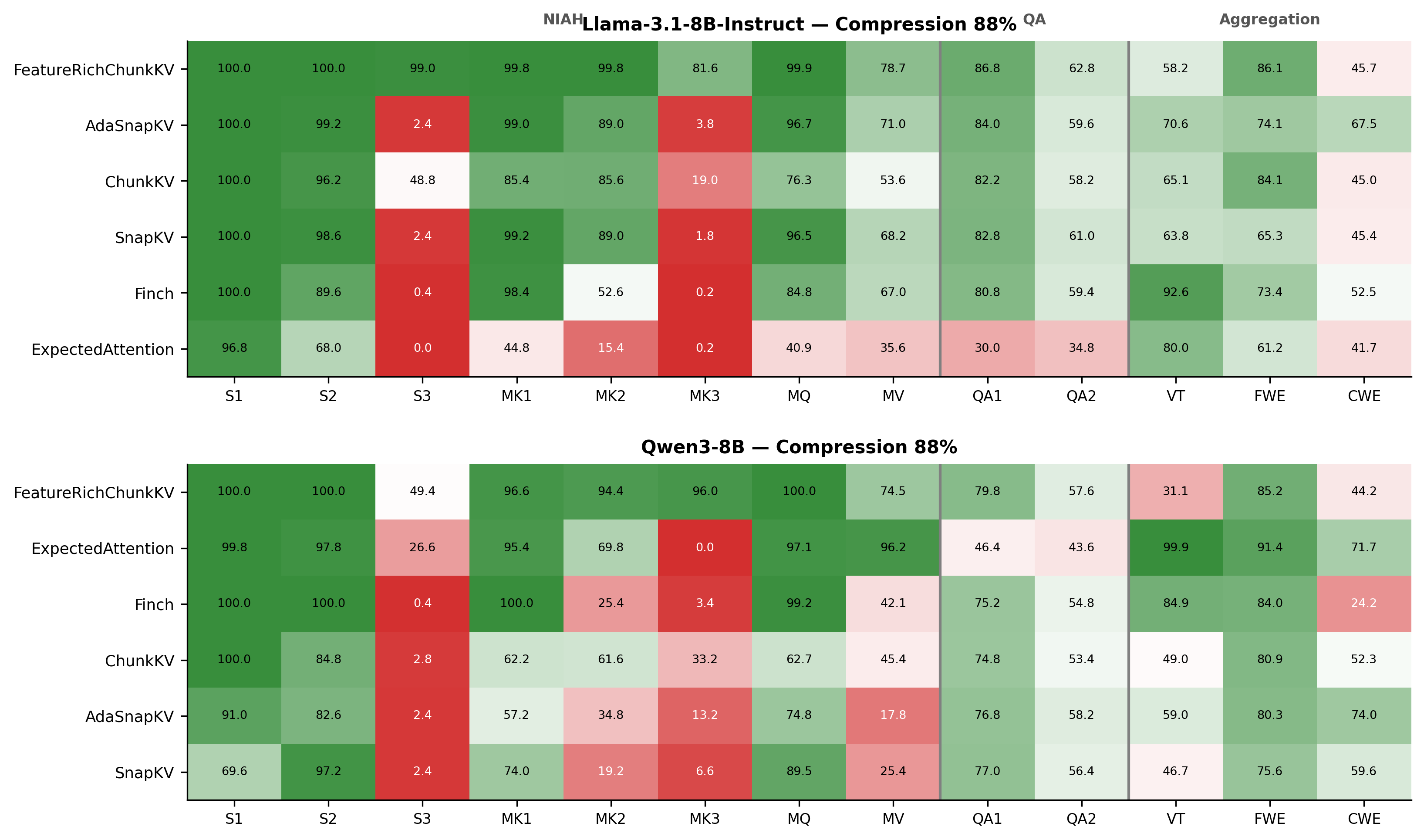}
\caption{Per-task RULER scores at $r{=}0.88$ and 4k context. Rows are KV-compression methods and columns are the 13 RULER subtasks: single-needle (S1--S3), multi-key (MK1--MK3), multi-query (MQ), multi-value (MV), QA (QA1--QA2), variable tracking (VT), frequent-word extraction (FWE), and common-word extraction (CWE). Top: Llama-3.1-8B-Instruct. Bottom: Qwen3-8B. Cell values are absolute accuracies; color encodes score magnitude rather than FRC-minus-baseline deltas.}
\label{fig:heatmap}
\end{figure}

The main text presents Figure~\ref{fig:delta_baseline} (drop relative to the no-compression upper bound across all cells), Figure~\ref{fig:heatmap} (per-subtask absolute scores at $r{=}0.88$), and Figure~\ref{fig:compression-overview} (high-compression slopegraph and rank trajectory). For completeness, Figures~\ref{fig:app_ruler4k} and~\ref{fig:app_ruler8k} report the same sweep data as conventional accuracy-versus-compression-ratio line plots at 4k and 8k, respectively.

\begin{table}[ht]
\caption{Per-task RULER scores at 4k context for Llama-3.1-8B-Instruct~\citep{grattafiori2024llama} across compression ratios $r \in \{0.25, 0.50, 0.75, 0.80, 0.88\}$. Columns group raw subtasks into single-needle (SN), multi-needle (MN), multi-key (MK), multi-value (MV), variable tracking (VT), common-word extraction (CWE), frequent-word extraction (FWE), and QA; AVG is the unweighted mean over the 13 raw RULER subtasks. Bold entries indicate the rank-1 compressed method at that ratio.}

\label{tab:app_llama4k_pertask}
\centering
\small
\setlength{\tabcolsep}{3.5pt}
\begin{tabularx}{\linewidth}{l l *{9}{Y}}
\toprule
$r$ & Method & SN & MN & MK & MV & VT & CWE & FWE & QA & AVG \\
\midrule
\multirow{7}{*}{0.25}
 & No Compression       & 100.0 & 99.9 & 99.9 & 99.9 & 99.9 & 99.6 & 95.0 & 75.5 & 95.8 \\
 & FRC (ours)           & \textbf{100.0} & \textbf{99.9} & \textbf{99.9} & \textbf{99.9} & \textbf{99.9} & 99.5 & 93.7 & \textbf{75.2} & \textbf{95.6} \\
 & ChunkKV~\citep{liu2025chunkkv}      & 100.0 & 99.7 & 99.7 & 99.5 & 99.9 & \textbf{99.6} & 93.4 & 75.1 & 95.5 \\
 & AdaSnapKV~\citep{feng2024ada}  & 98.9 & 99.9 & 99.8 & 99.9 & 99.9 & 99.5 & \textbf{94.7} & 75.3 & 95.4 \\
 & ExpectedAttention~\citep{devoto2025expected} & 99.8 & 99.8 & 99.7 & 99.6 & 99.8 & 99.7 & 94.7 & 73.0 & 95.2 \\
 & SnapKV~\citep{li2024snapkv}        & 96.1 & 99.8 & 99.5 & 99.7 & 99.2 & 99.4 & 93.6 & 75.0 & 94.5 \\
 & Finch~\citep{corallo2024finch}          & 99.9 & 99.8 & 92.1 & 98.8 & \textbf{100.0} & 99.3 & 94.1 & 75.2 & 93.7 \\
\midrule
\multirow{7}{*}{0.50}
 & No Compression       & 100.0 & 99.9 & 99.9 & 99.9 & 99.9 & 99.6 & 95.0 & 75.5 & 95.8 \\
 & FRC (ours)           & \textbf{100.0} & 99.8 & \textbf{99.9} & \textbf{99.9} & 99.8 & 98.0 & 93.1 & 74.7 & \textbf{95.4} \\
 & ChunkKV              & 99.9 & 99.5 & 98.3 & 98.2 & \textbf{99.9} & 97.5 & 91.9 & 75.0 & 94.7 \\
 & ExpectedAttention    & 97.7 & 99.4 & 92.5 & 97.8 & 99.5 & 99.5 & \textbf{94.9} & 69.8 & 92.4 \\
 & AdaSnapKV            & 82.9 & \textbf{99.9} & 98.7 & 99.3 & 99.4 & 99.2 & 93.4 & 75.0 & 91.2 \\
 & Finch                & 96.9 & 99.2 & 82.7 & 96.7 & 99.9 & 97.7 & 91.3 & \textbf{75.4} & 90.3 \\
 & SnapKV               & 79.5 & \textbf{99.9} & 94.7 & 99.3 & 97.6 & 97.7 & 89.9 & 74.9 & 89.0 \\
\midrule
\multirow{7}{*}{0.75}
 & No Compression       & 100.0 & 99.9 & 99.9 & 99.9 & 99.9 & 99.6 & 95.0 & 75.5 & 95.8 \\
 & FRC (ours)           & \textbf{100.0} & 99.8 & \textbf{99.7} & \textbf{99.7} & 97.4 & 86.9 & 89.7 & \textbf{75.0} & \textbf{94.0} \\
 & ChunkKV              & 99.3 & 97.9 & 90.7 & 94.0 & 97.9 & 83.1 & 89.4 & 74.4 & 90.8 \\
 & AdaSnapKV            & 68.5 & 99.7 & 81.3 & 97.0 & 94.8 & 95.0 & 87.7 & 74.9 & 82.5 \\
 & SnapKV               & 68.5 & \textbf{99.9} & 73.8 & 97.7 & 88.3 & 81.6 & 78.1 & 75.2 & 78.7 \\
 & Finch                & 74.4 & 96.5 & 60.4 & 90.0 & \textbf{99.6} & 86.4 & 85.3 & 73.0 & 77.6 \\
 & ExpectedAttention    & 69.8 & 94.8 & 67.7 & 84.2 & 96.7 & \textbf{95.6} & \textbf{93.7} & 54.5 & 75.9 \\
\midrule
\multirow{7}{*}{0.80}
 & No Compression       & 100.0 & 99.9 & 99.9 & 99.9 & 99.9 & 99.6 & 95.0 & 75.5 & 95.8 \\
 & FRC (ours)           & \textbf{100.0} & 99.8 & \textbf{99.6} & \textbf{99.6} & 90.2 & 79.1 & 88.7 & \textbf{75.6} & \textbf{92.9} \\
 & ChunkKV              & 98.6 & 96.4 & 85.9 & 91.2 & 94.2 & 74.4 & 89.1 & 73.9 & 88.2 \\
 & AdaSnapKV            & 67.5 & 99.5 & 75.1 & 95.2 & 92.0 & 90.2 & 83.0 & 74.0 & 79.7 \\
 & SnapKV               & 67.5 & 99.8 & 68.5 & 96.2 & 83.9 & 71.6 & 73.7 & 74.3 & 75.5 \\
 & Finch                & 68.3 & 93.9 & 56.7 & 84.2 & \textbf{99.4} & 78.2 & 81.1 & 72.0 & 73.5 \\
 & ExpectedAttention    & 65.8 & 88.8 & 53.3 & 79.0 & 92.2 & \textbf{88.1} & \textbf{92.3} & 46.5 & 68.5 \\
\midrule
\multirow{7}{*}{0.88}
 & No Compression       & 100.0 & 99.9 & 99.9 & 99.9 & 99.9 & 99.6 & 95.0 & 75.5 & 95.8 \\
 & FRC (ours)           & \textbf{99.7} & \textbf{99.9} & \textbf{93.7} & 78.7 & 58.2 & 45.7 & 86.1 & \textbf{74.8} & \textbf{84.5} \\
 & AdaSnapKV            & 67.2 & 96.7 & 63.9 & 71.0 & 70.6 & 67.5 & 74.1 & 71.8 & 70.5 \\
 & ChunkKV              & 81.7 & 76.3 & 63.3 & 53.6 & 65.1 & 45.0 & 84.1 & 70.2 & 69.2 \\
 & SnapKV               & 67.0 & 96.5 & 63.3 & 68.2 & 63.8 & 45.4 & 65.3 & 71.9 & 67.2 \\
 & Finch                & 63.3 & 84.8 & 50.4 & 67.0 & \textbf{92.6} & 52.5 & 73.4 & 70.1 & 65.5 \\
 & ExpectedAttention    & 54.9 & 40.9 & 20.1 & 35.6 & 80.0 & 41.7 & 61.2 & 32.4 & 42.3 \\
\bottomrule
\end{tabularx}
\end{table}

\footnotetext[1]{FRC weights $(0.55,\, 0.30,\, 0.15)$; ChunkKV chunk length $C{=}20$; baseline configs in Appendix~\ref{app:configs}.}
\footnotetext[2]{Bolded entries indicate rank-1 within method column at that ratio.}

\begin{table}[ht]
\caption{Per-task RULER scores at 4k context for Qwen3-8B~\citep{qwen3} across compression ratios $r \in \{0.25, 0.50, 0.75, 0.80, 0.88\}$. Columns and bolding follow Table~\ref{tab:app_llama4k_pertask}.}

\label{tab:app_qwen4k_pertask}
\centering
\small
\setlength{\tabcolsep}{3.5pt}
\begin{tabularx}{\linewidth}{l l *{9}{Y}}
\toprule
$r$ & Method & SN & MN & MK & MV & VT & CWE & FWE & QA & AVG \\
\midrule
\multirow{7}{*}{0.25}
 & No Compression       & 100.0 & 99.9 & 100.0 & 100.0 & 100.0 & 98.9 & 95.1 & 72.6 & 95.3 \\
 & AdaSnapKV            & \textbf{99.7} & \textbf{100.0} & \textbf{100.0} & 99.8 & \textbf{100.0} & 99.0 & 95.5 & \textbf{72.6} & \textbf{95.3} \\
 & ChunkKV              & \textbf{100.0} & 100.0 & 100.0 & 100.0 & \textbf{100.0} & \textbf{99.0} & 94.6 & 72.2 & 95.2 \\
 & Finch                & \textbf{100.0} & \textbf{100.0} & 99.7 & \textbf{100.0} & \textbf{100.0} & 98.5 & \textbf{95.7} & 72.0 & 95.2 \\
 & FRC (ours)           & \textbf{100.0} & \textbf{100.0} & \textbf{100.0} & \textbf{100.0} & \textbf{100.0} & 99.0 & 94.7 & 71.9 & 95.2 \\
 & ExpectedAttention    & \textbf{100.0} & \textbf{100.0} & 99.8 & \textbf{100.0} & \textbf{100.0} & 98.9 & 95.5 & 71.5 & 95.1 \\
 & SnapKV               & 98.9 & \textbf{100.0} & 98.9 & 99.7 & 99.4 & \textbf{99.0} & 95.3 & 72.5 & 94.7 \\
\midrule
\multirow{7}{*}{0.50}
 & No Compression       & 100.0 & 99.9 & 100.0 & 100.0 & 100.0 & 98.9 & 95.1 & 72.6 & 95.3 \\
 & FRC (ours)           & \textbf{100.0} & \textbf{100.0} & \textbf{99.9} & \textbf{100.0} & \textbf{100.0} & 98.1 & 93.1 & 71.9 & \textbf{95.0} \\
 & ExpectedAttention    & \textbf{100.0} & 100.0 & 99.7 & \textbf{100.0} & \textbf{100.0} & \textbf{99.1} & \textbf{95.4} & 69.5 & 94.8 \\
 & ChunkKV              & 98.9 & 100.0 & 98.7 & 97.5 & 100.0 & 98.1 & 92.1 & 72.2 & 94.2 \\
 & Finch                & 98.4 & \textbf{100.0} & 96.6 & 100.0 & 100.0 & 95.2 & 94.3 & 70.4 & 93.5 \\
 & AdaSnapKV            & 82.1 & 100.0 & 99.2 & 89.2 & 99.8 & 98.9 & 94.6 & 71.3 & 89.9 \\
 & SnapKV               & 80.3 & 100.0 & 92.9 & 87.0 & 97.0 & 98.6 & 92.9 & \textbf{72.0} & 87.6 \\
\midrule
\multirow{7}{*}{0.75}
 & No Compression       & 100.0 & 99.9 & 100.0 & 100.0 & 100.0 & 98.9 & 95.1 & 72.6 & 95.3 \\
 & FRC (ours)           & \textbf{99.9} & \textbf{100.0} & \textbf{99.9} & \textbf{99.7} & 97.8 & 91.0 & 90.2 & \textbf{71.6} & \textbf{93.9} \\
 & ExpectedAttention    & 95.0 & 100.0 & 83.5 & 99.8 & \textbf{100.0} & \textbf{97.6} & \textbf{94.1} & 60.1 & 88.3 \\
 & ChunkKV              & 90.9 & 97.9 & 87.3 & 84.7 & 97.9 & 87.4 & 87.7 & 70.7 & 87.0 \\
 & Finch                & 75.8 & 100.0 & 79.1 & 87.5 & 99.8 & 65.7 & 93.3 & 68.1 & 80.6 \\
 & AdaSnapKV            & 67.9 & 99.9 & 84.5 & 42.1 & 94.4 & 96.5 & 89.7 & 70.8 & 78.6 \\
 & SnapKV               & 67.7 & 100.0 & 69.1 & 49.6 & 85.1 & 92.6 & 85.3 & 71.0 & 74.3 \\
\midrule
\multirow{7}{*}{0.80}
 & No Compression       & 100.0 & 99.9 & 100.0 & 100.0 & 100.0 & 98.9 & 95.1 & 72.6 & 95.3 \\
 & FRC (ours)           & \textbf{99.9} & \textbf{100.0} & \textbf{99.8} & 98.0 & 84.2 & 82.6 & 89.1 & \textbf{70.9} & \textbf{91.9} \\
 & ExpectedAttention    & 88.7 & 99.8 & 75.3 & \textbf{99.6} & \textbf{100.0} & 94.0 & 93.6 & 54.1 & 83.6 \\
 & ChunkKV              & 85.3 & 92.2 & 74.9 & 76.8 & 93.2 & 79.1 & 86.5 & 69.8 & 81.7 \\
 & AdaSnapKV            & 67.5 & 99.4 & 71.9 & 33.6 & 88.0 & \textbf{94.2} & 86.7 & 70.3 & 73.9 \\
 & Finch                & 70.3 & 100.0 & 63.8 & 75.0 & 98.4 & 53.0 & \textbf{90.9} & 67.5 & 73.4 \\
 & SnapKV               & 66.0 & 99.8 & 57.9 & 40.4 & 75.9 & 86.0 & 83.1 & 70.5 & 69.1 \\
\midrule
\multirow{7}{*}{0.88}
 & No Compression       & 100.0 & 99.9 & 100.0 & 100.0 & 100.0 & 98.9 & 95.1 & 72.6 & 95.3 \\
 & FRC (ours)           & \textbf{83.1} & \textbf{100.0} & \textbf{95.7} & 74.5 & 31.1 & 44.2 & 85.2 & \textbf{68.7} & \textbf{77.6} \\
 & ExpectedAttention    & 74.7 & 97.1 & 55.1 & \textbf{96.2} & \textbf{99.9} & 71.7 & \textbf{91.4} & 45.0 & 72.0 \\
 & Finch                & 66.8 & 99.2 & 42.9 & 42.1 & 84.9 & 24.2 & 84.0 & 65.0 & 61.0 \\
 & ChunkKV              & 62.5 & 62.7 & 52.3 & 45.4 & 49.0 & 52.3 & 80.9 & 64.1 & 58.7 \\
 & AdaSnapKV            & 58.7 & 74.8 & 35.1 & 17.8 & 59.0 & 74.0 & 80.3 & 67.5 & 55.5 \\
 & SnapKV               & 56.4 & 89.5 & 33.3 & 25.4 & 46.7 & 59.6 & 75.6 & 66.7 & 53.8 \\
\bottomrule
\end{tabularx}
\end{table}

\footnotetext[3]{FRC weights $(0.55,\, 0.30,\, 0.15)$; ChunkKV chunk length $C{=}20$; baseline configs in Appendix~\ref{app:configs}.}
\footnotetext[4]{Bolded entries indicate rank-1 within method column at that ratio.}

\begin{table}[ht]
\caption{ RULER 8k aggregate accuracy for Llama-3.1-8B-Instruct~\citep{grattafiori2024llama} across compression ratios. AVG is the unweighted arithmetic mean over the 13 raw RULER subtasks emitted by the evaluation harness enumerated in Section~\ref{sec:setup}. Bold entries indicate the rank-1 method at that ratio. The per-task 8k breakdown is released as an auxiliary artifact alongside the paper; we report the aggregate here because it is what the headline statements in Section~\ref{sec:results} use.\protect\footnotemark[5]\protect\footnotemark[6]}
\label{tab:app_llama8k}
\centering
\small
\begin{tabularx}{\linewidth}{l *{5}{Y}}
\toprule
Method & $r{=}0.25$ & $r{=}0.50$ & $r{=}0.75$ & $r{=}0.80$ & $r{=}0.88$ \\
\midrule
No Compression                               & 94.6 & 94.6 & 94.6 & 94.6 & 94.6 \\
FRC (ours)                                   & 94.4 & \textbf{94.1} & \textbf{92.9} & \textbf{92.2} & \textbf{89.7} \\
ChunkKV~\citep{liu2025chunkkv}                      & \textbf{94.5} & \textbf{94.1} & 91.4 & 89.7 & 83.5 \\
AdaSnapKV~\citep{feng2024ada}                  & 94.4 & 92.7 & 83.5 & 80.7 & 73.3 \\
SnapKV~\citep{li2024snapkv}                        & 93.8 & 89.2 & 79.9 & 76.9 & 70.0 \\
Finch~\citep{corallo2024finch}                          & 92.7 & 90.4 & 79.6 & 76.1 & 69.3 \\
ExpectedAttention~\citep{devoto2025expected}       & 93.8 & 89.1 & 71.6 & 65.2 & 37.9 \\
\bottomrule
\end{tabularx}
\end{table}

\footnotetext[5]{FRC weights $(0.55,\, 0.30,\, 0.15)$; ChunkKV chunk length $C{=}20$; baseline configs in Appendix~\ref{app:configs}.}
\footnotetext[6]{Bolded entries indicate rank-1 within method column at that ratio.}

\begin{table}[ht]
\caption{ RULER 8k aggregate accuracy for Qwen3-8B~\citep{qwen3} across compression ratios. Conventions follow Table~\ref{tab:app_llama8k}. Note the cross-model reordering: the strongest baseline at $r{=}0.88$ on Qwen-8k is ChunkKV, against which FRC gains $+13.9$; on Llama-8k FRC gains $+6.2$ over the same baseline. Per-task 8k scores are released as an auxiliary artifact.\protect\footnotemark[7]\protect\footnotemark[8]}
\label{tab:app_qwen8k}
\centering
\small
\begin{tabularx}{\linewidth}{l *{5}{Y}}
\toprule
Method & $r{=}0.25$ & $r{=}0.50$ & $r{=}0.75$ & $r{=}0.80$ & $r{=}0.88$ \\
\midrule
No Compression                               & 93.9 & 93.9 & 93.9 & 93.9 & 93.9 \\
FRC (ours)                                   & 93.7 & 93.2 & \textbf{91.9} & \textbf{91.3} & \textbf{88.6} \\
ChunkKV~\citep{liu2025chunkkv}                      & 93.7 & 92.9 & 87.6 & 84.0 & 74.7 \\
AdaSnapKV~\citep{feng2024ada}                  & 93.8 & 88.3 & 80.3 & 77.3 & 68.0 \\
SnapKV~\citep{li2024snapkv}                        & 92.6 & 86.6 & 78.0 & 74.7 & 64.9 \\
Finch~\citep{corallo2024finch}                          & 92.6 & 90.5 & 85.6 & 79.9 & 70.2 \\
ExpectedAttention~\citep{devoto2025expected}       & \textbf{93.9} & \textbf{93.5} & 87.3 & 82.3 & 69.9 \\
\bottomrule
\end{tabularx}
\end{table}

\footnotetext[7]{FRC weights $(0.55,\, 0.30,\, 0.15)$; ChunkKV chunk length $C{=}20$; baseline configs in Appendix~\ref{app:configs}.}
\footnotetext[8]{Bolded entries indicate rank-1 within method column at that ratio.}

\section{Per-task heatmaps}
\label{app:heatmap}

This appendix provides per-task heatmaps at the most aggressive evaluated compression ratio, $r{=}0.88$, complementing the aggregate tables in Appendix~\ref{app:full_tables}. Each cell reports the absolute RULER subtask score for one method and subtask. The panels show Llama-3.1-8B-Instruct and Qwen3-8B at 4k context; darker green indicates higher accuracy and darker red indicates lower accuracy.

\section{OpenEvolve configurations}
\label{app:configs}

\noindent\textbf{Mutator and infrastructure.}
All evolution runs use OpenEvolve~v0.5.1 with the \texttt{qwen3.5:35b-a3b-q4\_K\_M} mutator served via a local Ollama instance. The scaled run (\texttt{B-redux-v3b}) uses the no-reasoning variant of the same checkpoint to keep per-iteration latency tractable at 1{,}500 iterations. Each island maintains a MAP-Elites grid over two behavioral axes: program complexity (bucketed source length) and code-embedding diversity (cosine distance in a fixed embedding space over the editable region). Parent sampling is a softmax over the utility $0.7\cdot\text{fitness} + 0.3\cdot\text{recency}$, so recent improvements remain reachable as parents without displacing high-fitness elites. Islands exchange elites under a ring topology with migration rate $\rho_{\text{mig}}$ every $\tau_{\text{mig}}$ generations, with both values tabulated per run in the supplemental YAML configs.

\noindent\textbf{Cascade evaluator.}
The cascade evaluator runs three stages. Stage-0 enforces syntax and import gates, rejecting malformed candidates without scoring. Stage-1 runs the candidate on a \texttt{small\_fraction}${=}0.02$ subsample of the RULER reward harness and gates at quality 45.0. Stage-2 promotes survivors to a \texttt{full\_fraction}${=}0.05$ evaluation; only stage-2 survivors enter the archive.
\begin{table}[ht]
\caption{Per-iteration hyperparameters for evolution runs referenced in the paper. \texttt{V1}--\texttt{V5} trace the discovery path to FRC; \texttt{B-Redux} runs are diversity-controlled family-crossing runs; \texttt{v3a}--\texttt{v3c} are the older scaled-redux engineering runs.}
\label{tab:cfg-evolve}
\centering
\footnotesize
\setlength{\tabcolsep}{3.2pt}
\begin{tabularx}{\linewidth}{lrrrrr l X}
\toprule
Run & iter & pop & islands & RNG & $r_{\text{search}}$ & mutator & notes \\
\midrule
V1 & 100 & 200 & 3 & 42 & 0.12 & reasoning & tokenwise top-$k$ harness \\
V2 & 100 & 200 & 3 & 42 & 0.12 & reasoning & re-seeded with V1 best \\
V3 & 100 & 200 & 3 & 42 & 0.12 & reasoning & re-seeded with V2 best \\
V4 & 100 & 200 & 3 & 42 & 0.12 & reasoning & query-aware extension \\
V5 & 100 & 200 & 3 & 42 & 0.83 & reasoning & ChunkKV-style selection; literature-audited seed \\
\midrule
B-Redux-1 & 500 & 400 & 9 & 4242 & 0.83 & no-reason, T=0.7 & V1 5-signal seed \\
B-Redux-2 & 500 & 400 & 9 & 4242 & 0.83 & no-reason, T=1.0 & reward-hacked; fixed \\
B-Redux-3v1 & 500 & 400 & 9 & 4242 & 0.83 & no-reason, T=1.2 & too noisy \\
B-Redux-3v2 & 500 & 600 & 12 & 4242 & 0.83 & no-reason, T=0.9 & reward-hacked; fixed \\
\midrule
v3b & 1500 & 600 & 12 & 1337 & 0.83 & no-reason & iter-1073 milestone \\
v3a & 1500 & 600 & 12 & 42 & 0.83 & no-reason & stalled before checkpoint \\
v3c & 1500 & 600 & 12 & 2718 & 0.83 & no-reason & stalled before checkpoint \\
\bottomrule
\end{tabularx}
\end{table}

\noindent\textbf{Editable contract evolution.}
The editable contract (the slice of the policy module that OpenEvolve may mutate) went through three revisions. The \emph{v1} contract was a roughly 30-line policy module with two entry points (\texttt{score\_tokens} and \texttt{select\_tokens\_to\_keep}); enforcement was loose, with no type discipline on \texttt{score\_tokens}'s return value and no check that the selected token set matched the budget. The \emph{v2} contract tightens the post-condition invariants on \texttt{select\_tokens\_to\_keep} (the selected count must equal the budget, indices must be unique, and every index must lie in $[0, L)$) and is the contract used for V1 through V5 and the entire scaled-redux-v3 series. The \emph{v3} contract is a post-mortem revision drafted after the iteration-1264 reward hack in \texttt{B-redux-v3b}: it tightens the \texttt{EVOLVE-BLOCK} boundaries to close the diff-applier contract gap (OpenEvolve Issue~\#281, PRs~\#352 and~\#354), so that mutations cannot rewrite \texttt{select\_tokens\_to\_keep} or other locked symbols outside the intended editable region. The v3 contract is not used for any results reported in this paper, but it informs the future-work discussion in Section~\ref{sec:limits} and Appendix~\ref{app:redux}.

\noindent\textbf{Configuration files and reproducibility.}
The complete YAML configurations for every row of Table~\ref{tab:cfg-evolve} are included as \texttt{redux/configs/*.yaml} in the supplemental materials. Each file pins the mutator checkpoint, cascade fractions, island and migration parameters, and editable-contract version. Every reported result corresponds to a checkpointed run with a logged seed, a configuration hash over the YAML, and a code hash over the policy module and evaluator; running the supplemental harness against a checkpoint regenerates the cascade-stage scores reported in this paper.

\subsection{Reproducibility checklist}
\label{app:reprochecklist}

We pin the following components and provide canonical command lines for the four headline results.

\noindent\textbf{Versions and infrastructure.}
\begin{itemize}\setlength\itemsep{2pt}
\item \textbf{Models:} \texttt{meta-llama/Llama-3.1-8B-Instruct}~\citep{grattafiori2024llama}, \texttt{Qwen/Qwen3-8B}~\citep{qwen3}.
\item \textbf{KVPress:} pinned at the commit shipped in the supplemental materials (\texttt{kvpress/} directory). Baselines (SnapKV, ChunkKV, AdaSnapKV, Finch, ExpectedAttention) run through the official \texttt{evaluation/evaluate.py} on this commit.
\item \textbf{OpenEvolve:} v0.5.1, used via \texttt{openevolve.api.run\_evolution}.
\item \textbf{Mutator LLM:} \texttt{qwen3.5:35b-a3b-q4\_K\_M} served through Ollama on a local network endpoint; reasoning variant for V1 through V5, no-reasoning variant for the scaled redux runs.
\item \textbf{Hardware:} NVIDIA H200 GPUs (single-GPU per evaluation run; 1 to 3 GPUs per parallel evolution campaign).
\end{itemize}

\noindent\textbf{Seeds and fractions.}
RNG seeds: $42$ for V1 through V5; $4242$ for B-Redux-1 through B-Redux-3v2; $1337$ for v3b. Cascade fractions: \texttt{small\_fraction}${=}0.02$ for Stage-1 (gated at $\geq 45.0$), \texttt{full\_fraction}${=}0.10$ for Stage-2 except B-Redux which used $0.05$, with combined fitness $0.25\,S_1 + 0.75\,S_2$. All headline results in Figure~\ref{fig:heatmap_r088} and Tables~\ref{tab:ablation}, \ref{tab:within_frc}, \ref{tab:redux} are full-RULER aggregates (\texttt{fraction}${=}1.0$) on the held-out evaluation grid.

\noindent\textbf{Canonical command lines.} The four headline results are regenerated by:
\begin{itemize}\setlength\itemsep{2pt}
\item \emph{Cross-model FRC evaluation} (Figure~\ref{fig:heatmap_r088}): \texttt{benchmark\_vs\_official.py --policy-path frc\_v5.py --model \{llama,qwen\} --dataset ruler --data-dir \{4096,8192\} --compression-ratio \{0.25,0.50,0.75,0.80,0.88\} --fraction 1.0 --evolved-query-aware}.
\item \emph{Scorer-versus-structure ablation} (Table~\ref{tab:ablation}): the same harness with \texttt{--policy-path \{v1\_tokenwise.py, frc\_chunk.py, frc\_tokenwise.py\}} at \texttt{--compression-ratio 0.88}.
\item \emph{Within-FRC ablation} (Table~\ref{tab:within_frc}): the same harness with \texttt{--policy-path ablations/\{full\_evolved, no\_max\_head, no\_neighbor, local\_only\}.py} at \texttt{--compression-ratio 0.83}.
\item \emph{B-Redux family-crossing} (Table~\ref{tab:redux}): \texttt{run\_b\_redux\_v\{1,2,3v1,3v2\}.py} with \texttt{REDUX\_TEMPERATURE=\{0.7,1.0,1.2,0.9\} REDUX\_NUM\_ISLANDS=\{9,9,9,12\}} from \texttt{prefill\_openevolve/seeds/v1\_rich\_features\_seed.py}.
\end{itemize}
The supplemental release contains all four launchers, the seed and ablation policy modules, and the YAML config files; an anonymized version is provided as supplemental materials, and the public repository will be released upon deanonymization.

\section{Editable feature set}
\label{app:features}

This appendix specifies the editable feature set referenced in Section~\ref{sec:method} and Section~\ref{sec:formulation}.

\paragraph{Interface contract.}
Every evolved policy is a Python module that operates against a single frozen data structure, \texttt{PrefillContext}, exposing \textbf{17 token-level signals} (each a NumPy array of shape $[L]$, min-max normalized) and \textbf{5 scalar metadata fields}. The module must export exactly two entry points: \texttt{score\_tokens(ctx) -> ndarray[L]}, returning a per-token saliency score, and \texttt{select\_tokens\_to\_keep(ctx) -> ndarray[budget]}, returning the indices retained under the cache budget. The mutable surface is delimited by \texttt{EVOLVE-BLOCK} markers: OpenEvolve may rewrite anything between them, while the surrounding harness (signature, imports, output-shape contracts, and the selection wrapper used in the constrained-redux runs of \S\ref{sec:analysis}) is frozen. Table~\ref{tab:features} enumerates the 17 signals and their literature provenance; the three signals used by the FRC scorer of Eq.~\ref{eq:score} are flagged in the final column.

\begin{table}[ht]
\caption{Signals exposed to evolved policies via \texttt{PrefillContext}. Continuous attention-, key-, and chunk-mean features are MaxAbs-normalized before scoring: nonzero vectors are scaled by their maximum absolute value. Position features, rank features, boolean masks, and scalar metadata remain in their native bounded scales. For GQA models, attention-derived signals are computed from the KV-head/group proxy $\widetilde A$ in Listing~\ref{alg:proxy}. The column \emph{FRC?} marks the three signals used by the final FRC scorer.}
\label{tab:features}
\centering
\footnotesize
\begin{tabular}{llp{5.0cm}lc}
\toprule
Signal & Family & Definition (concise) & Scale & FRC? \\
\midrule
\multicolumn{5}{l}{\emph{Attention statistics}} \\
\texttt{attn\_received} &
global &
$\tfrac{1}{H_{\mathrm{kv}}Q}\sum_{g,q}\widetilde A_{g,q,t}$ &
MaxAbs & \\

\texttt{tail\_attn\_received} &
local &
$\tfrac{1}{H_{\mathrm{kv}}W}\sum_g\sum_{q=Q-W}^{Q-1}\widetilde A_{g,q,t}$ &
MaxAbs & \\

\texttt{local\_attn\_received} &
local &
Alias/copy of \texttt{tail\_attn\_received}; local tail-window attention, $W{=}32$ &
MaxAbs & $\star$ \\

\texttt{global\_attn\_received} &
global &
Mean attention before the tail window; falls back to \texttt{attn\_received} if no prefix exists &
MaxAbs & \\

\texttt{max\_head\_attn\_received} &
head &
$\max_g \tfrac{1}{Q}\sum_q \widetilde A_{g,q,t}$, maximum over KV-group-averaged heads &
MaxAbs & $\star$ \\

\texttt{head\_consistency} &
head &
$1/(1+\mathrm{Var}_g(\tfrac{1}{Q}\sum_q \widetilde A_{g,q,t}))$ &
MaxAbs & \\

\texttt{attn\_concentration} &
head &
$1-\mathrm{entropy}(\text{head shares})/\log(\max(H_{\mathrm{kv}},2))$ &
MaxAbs & \\

\texttt{neighbor\_attn\_density} &
local &
Moving average of \texttt{attn\_received} over $W_d{=}\max(3,\min(33,2\lfloor L/64\rfloor+1))$ &
MaxAbs & $\star$ \\

\midrule
\multicolumn{5}{l}{\emph{Key-space and chunk features}} \\
\texttt{key\_norms} &
key &
Mean headwise key norm $\lVert k_t\rVert_2$ &
MaxAbs & \\

\texttt{key\_change\_norms} &
key &
Mean headwise $\lVert k_t-k_{t-1}\rVert_2$, prepending zero &
MaxAbs & \\

\texttt{chunk\_mean\_attn} &
chunk &
Mean \texttt{attn\_received} within the token's coarse chunk &
MaxAbs & \\

\texttt{chunk\_rank} &
chunk &
Normalized rank of the token's chunk by mean attention &
native $[0,1]$ & \\

\midrule
\multicolumn{5}{l}{\emph{Position, masks, and scalar metadata}} \\
\texttt{positions} &
pos &
Absolute token index $t$ &
native & \\

\texttt{normalized\_positions} &
pos &
$t/\max(L-1,1)$ &
native $[0,1]$ & \\

\texttt{distance\_to\_end} &
pos &
$(L-1-t)/\max(L-1,1)$ &
native $[0,1]$ & \\

\texttt{sink\_mask} &
mask &
Boolean mask for the first $s$ sink tokens &
native bool & \\

\texttt{is\_tail\_mask} &
mask &
Boolean mask for tokens in the local tail window &
native bool & \\

\texttt{cache\_budget} &
scalar &
$\max(1,\lfloor L(1-r)\rfloor)$ for tokenwise policies &
native int & \\

\texttt{compression\_ratio} &
scalar &
Discarded KV fraction $r$ &
native float & \\

\texttt{num\_heads}, \texttt{q\_len}, \texttt{kv\_len} &
scalar &
Number of proxy heads, query length, and KV length &
native int & \\
\bottomrule
\end{tabular}
\end{table}

\paragraph{Scalar metadata.}
Alongside the 17 token-level arrays, \texttt{PrefillContext} exposes five scalar fields that policies use for adaptive parameterization: sequence length $L$, retained budget $K$, head count $H$, query count $Q$, and layer index $\ell$. These fields are passed through unmodified because most consumers need raw scale rather than a normalized variant: for example, the width-adaptive density window $W_d$ depends on $L$, chunking heuristics depend on $K$ and $L$, and per-layer reweighting uses $\ell$. Policies in our search did not rewrite these fields; they only read them.

\paragraph{Normalization protocol and output invariants.}
Each of the 17 token-level arrays is min-max normalized at construction, so any linear combination written by an evolved \texttt{score\_tokens} is well-scaled regardless of which signals are mixed and at what weights. Scalar metadata is used as-is. At the evaluation boundary we enforce three output invariants on \texttt{select\_tokens\_to\_keep}: the returned index array contains exactly $K$ entries, entries are unique (no duplicates), and every entry lies in $[0, L)$. Programs that violate any invariant are rejected by the cascade evaluator before scoring, so OpenEvolve sees only policies that produce valid retained-token sets.

\section{Latency and compute}
\label{app:latency}

This appendix details the latency claim referenced in Section~\ref{sec:method}.

\paragraph{Inference latency.}
FRC adds negligible per-token cost during prefill: three element-wise statistics over the attention tensor, a linear combination, and a top-$k$ selection. The method is single-pass (no auxiliary forward passes are needed), so the only additional work relative to a no-compression baseline is feature computation followed by index selection, both performed once per layer at the end of prefill. Per-prefill overhead is $\mathcal{O}(LH)$ for \texttt{local\_attn\_received}, $\mathcal{O}(L)$ for \texttt{neighbor\_attn\_density} (a 1D average pool over column-summed attention), $\mathcal{O}(LH)$ for \texttt{max\_head\_attn\_received}, and $\mathcal{O}(L \log L)$ for top-$k$ selection. All four terms are dominated by the $\mathcal{O}(L^2 H)$ cost of attention itself. Table~\ref{tab:latency} reports wall-clock prefill latency on Llama-3.1-8B-Instruct on a single NVIDIA H100 80GB: a complete forward pass with the press active (prefill plus compress), mean $\pm$ standard deviation over 100 timed iterations after 10 warmup runs (single batch). The original NumPy-based output of evolutionary search (\textbf{FRC-CPU}) requires a GPU-to-CPU transfer for feature computation; our GPU-native PyTorch reimplementation (\textbf{FRC-GPU}, using \texttt{torch.mean}, \texttt{torch.max}, \texttt{F.avg\_pool1d}, and \texttt{scatter\_add\_}) preserves the scoring exactly while eliminating the transfer. FRC-GPU is 1.1 to 1.6$\times$ faster than ChunkKV across all configurations and 1.1 to 1.3$\times$ slower than SnapKV alone, reflecting the cost of computing three signals instead of one.

\begin{table}[ht]
\caption{Prefill latency (ms) on Llama-3.1-8B-Instruct (NVIDIA H100 80GB). Lower is better. Mean $\pm$ std over 100 timed iterations after 10 warmup runs, single batch. Boldface marks the faster method among FRC-GPU and ChunkKV at each operating point; SnapKV is shown as a single-signal speed reference rather than a quality-matched baseline.}
\label{tab:latency}
\centering
\small
\begin{tabular}{llrrrr}
\toprule
Context & CR & FRC-GPU (ours) & FRC-CPU & ChunkKV & SnapKV \\
\midrule
\multirow{3}{*}{$L{=}4096$}
 & 50\% & \textbf{204.5} $\pm$ 1.6  & 256.2 $\pm$ 35.3  & 313.2 $\pm$ 0.7 & 143.5 $\pm$ 0.9 \\
 & 83\% & \textbf{170.0} $\pm$ 0.5  & 239.8 $\pm$ 29.8  & 199.1 $\pm$ 0.4 & 144.2 $\pm$ 1.0 \\
 & 88\% & \textbf{168.8} $\pm$ 0.4  & 266.2 $\pm$ 13.3  & 187.1 $\pm$ 0.5 & 146.4 $\pm$ 0.6 \\
\midrule
\multirow{3}{*}{$L{=}8192$}
 & 50\% & \textbf{407.5} $\pm$ 0.9  & 797.6 $\pm$ 104.0 & 632.1 $\pm$ 0.5 & 304.1 $\pm$ 2.3 \\
 & 83\% & \textbf{341.0} $\pm$ 0.7  & 860.1 $\pm$ 18.8  & 404.6 $\pm$ 1.0 & 303.0 $\pm$ 2.4 \\
 & 88\% & \textbf{331.4} $\pm$ 1.2  & 941.9 $\pm$ 109.4 & 372.1 $\pm$ 0.6 & 303.7 $\pm$ 2.3 \\
\bottomrule
\end{tabular}
\end{table}

\paragraph{Search compute (OpenEvolve).}
Search-time cost is dominated by two stages: (i) mutator generation by a local Ollama instance running \texttt{qwen3.5:35b-a3b-q4\_K\_M} ($\approx$30s per proposed diff), and (ii) cascade evaluation, where Stage-1 (smaller pool, $\approx$60 to 90s) gates entry to Stage-2 (full pool, $\approx$300 to 450s). Only $\approx$12\% of generated programs survive Stage-1. The short V1 through V5 runs (100 iterations, population 200, 3 islands) consume $\approx$4 to 6 hours wall-clock per run on a single H100 paired with the Ollama-host machine. The scaled-redux v3b run (1500 iterations, population 600, 12 islands) takes $\approx$3 weeks wall-clock end-to-end. Cascade gating is the main compute saver: the $\approx$88\% of programs filtered at Stage-1 never incur the $5\times$ longer Stage-2 cost. Memory savings at inference are substantial: at $r{=}0.75$ on an $L{=}8192$ context with $H{=}32$ heads, $d{=}128$, and 32 layers in fp16, the KV cache shrinks from $\approx$4~GB to $\approx$1~GB.

\section{LongBench-v2 detail}
\label{app:longbench}

This appendix expands the LongBench-v2 result mentioned in Section~\ref{sec:limits}. The headline numbers reproduce the main-text result; the appendix adds (i) the per-subset breakdown that locates FRC underperformance in long-context multi-hop comprehension, (ii) the Qwen capability-ceiling diagnosis that explains why the Qwen LongBench-v2 column was excluded, and (iii) a null result from a LongBench-targeted evolution run, which we read as a further diagnostic trigger.

The setup mirrors Section~\ref{sec:setup}: zero-shot, $r{=}0.83$ compression, KVPress single-pass prefill-stage baselines, evaluated on Llama-3.1-8B-Instruct.

\paragraph{Llama-3.1-8B-Instruct results.} Table~\ref{tab:longbench_llama} reports per-method LongBench-v2 accuracy. ExpectedAttention is the strongest compressed method, retaining $91.7\%$ of the no-compression score; FRC retains $65.0\%$. The gap concentrates in the \emph{long} subset (passages over 32k tokens), where ExpectedAttention scores $0.176$ but every windowed scorer (FRC, ChunkKV, SnapKV, AdaSnapKV) scores exactly $0.000$. ExpectedAttention's key-norm estimator preserves uniform coverage across the full context, whereas windowed scorers concentrate retention near recent positions. That bias aligns with RULER's needle placement but discards early tokens that are critical for multi-hop comprehension on LongBench-v2.

\begin{table}[ht]
\centering
\small
\caption{LongBench-v2 accuracy (zero-shot, $r{=}0.83$, Llama-3.1-8B-Instruct). Higher is better. Finch did not produce metrics in our local environment.}
\label{tab:longbench_llama}
\begin{tabular}{lr}
\toprule
Method & Accuracy \\
\midrule
No Compression & 0.288 \\
ExpectedAttention & \textbf{0.264} \\
FRC (ours) & 0.187 \\
AdaSnapKV & 0.183 \\
ChunkKV & 0.183 \\
SnapKV & 0.181 \\
\bottomrule
\end{tabular}
\end{table}

\paragraph{Qwen3-8B sits at the base-model ceiling.} On Qwen, all six methods cluster within $0.004$ of one another (range $0.177$ to $0.181$), and every method scores $0.000$ on the long subset, including the no-compression baseline. The base model itself reaches only $0.181$, leaving no signal above the random-guess floor (LongBench-v2 is four-way multiple choice with base rate $0.25$). Compression cannot preserve information the base model did not retrieve in the first place, so per-method rankings here are uninformative. We therefore exclude Qwen3-8B from the LongBench-v2 headline summary and treat the result as a base-model capability ceiling rather than a property of any compression method.

\paragraph{A fourth diagnostic trigger.} Where fixed FRC underperforms, the framework's prescription is to re-run the discovery loop with a benchmark-appropriate cascade evaluator rather than to further hand-tune the same retrieval-oriented weights. We executed a 500-iteration LongBench-v2-driven evolution from the FRC seed; the run did not produce a competitive policy, with the tracked best remaining the initial candidate at stage-1/stage-2 = $10.0/8.0$ (combined $0.085$). This null result is a fourth diagnostic trigger, additional to the three identified in Section~\ref{sec:analysis} and currently unresolved: the FRC editable interface, scoped from a literature audit oriented to RULER-style retrieval, does not contain signals that the cascade can reward on LongBench-v2. LongBench-v2 requires qualitatively different retention signals (uniform key-norm coverage and document-section-boundary cues are plausible candidates), not a small reweighting of local-density features. The same recipe applied to RULER in Section~\ref{sec:formulation} (a literature-informed seed plus an interface scoped to the benchmark's structure) is what we expect to produce a competitive LongBench-v2 policy; we have not yet run it.

\section{Additional benchmarks}
\label{app:additional}

This appendix details the additional benchmarks referenced in Section~\ref{sec:limits}.

We use these benchmarks as an audit of the current FRC weights rather than as headline evidence. For each new benchmark, we first test fixed FRC directly: if it matches or beats the strongest KVPress baseline, we report the result; if it underperforms, we treat the benchmark as evidence motivating benchmark-specific evolution (Section~\ref{sec:analysis}).

\subsection{Needle-in-a-Haystack at 16k}
\label{app:niah}

Table~\ref{tab:niah16k} reports Needle-in-a-Haystack \citep{niah16k} results at $16$k context on Llama-3.1-8B-Instruct. The query-aware setting is largely saturated: most attention-window methods cluster near the no-compression score, and Finch is slightly stronger at every ratio. The no-query setting is more discriminative. There, FRC improves over ChunkKV at high compression (especially at $r{=}0.75$ and $r{=}0.88$), but it is not uniformly best: SnapKV and AdaSnapKV remain competitive at the deepest setting. We therefore treat NIAH 16k as supporting evidence that FRC's density signal helps aggressive retrieval compression, not as an independent headline result.

\begin{table}[ht]
\caption{NIAH 16k ROUGE-L F-score (Llama-3.1-8B-Instruct). Higher is better. Query-aware results are mostly saturated; no-query results are more discriminative.}
\label{tab:niah16k}
\centering
\small
\begin{tabular}{llrrr}
\toprule
Setting & Method & 50\% & 75\% & 88\% \\
\midrule
\multirow{6}{*}{Query-aware}
 & No Compression & 0.710 & 0.710 & 0.710 \\
 & FRC & 0.718 & 0.718 & 0.718 \\
 & ChunkKV & 0.718 & 0.718 & 0.718 \\
 & SnapKV & 0.718 & 0.718 & 0.710 \\
 & AdaSnapKV & 0.718 & 0.718 & 0.710 \\
 & Finch & \textbf{0.742} & \textbf{0.742} & \textbf{0.750} \\
\midrule
\multirow{6}{*}{No-query}
 & No Compression & 0.710 & 0.710 & 0.710 \\
 & FRC & 0.726 & 0.716 & 0.449 \\
 & ChunkKV & 0.718 & 0.680 & 0.321 \\
 & SnapKV & 0.742 & \textbf{0.750} & \textbf{0.612} \\
 & AdaSnapKV & 0.726 & \textbf{0.750} & 0.588 \\
 & ExpectedAttention & \textbf{0.750} & 0.492 & 0.194 \\
\bottomrule
\end{tabular}
\end{table}

Qwen3-8B at $16$k context exceeded our local memory budget under the KVPress evaluation harness; we therefore restrict NIAH 16k to Llama-3.1-8B-Instruct.

\subsection{InfiniteBench audit}
\label{app:infbench}

InfiniteBench results are mixed and do not support adding FRC as a main-paper claim. On \texttt{longbook\_choice\_eng}, ExpectedAttention is substantially stronger than FRC, suggesting that uniform or key-norm-style coverage is better for long-document comprehension. On \texttt{passkey} and \texttt{number\_string}, scores are nearly unchanged across methods and match the no-compression baseline, making the setting uninformative. On \texttt{kv\_retrieval}, most compressed methods collapse on the sampled subset. We include this audit to delimit FRC's operating regime and leave InfiniteBench-specific evolution to future work.
The unexpectedly low no-compression scores on \texttt{passkey} and \texttt{number\_string} indicate that this audit may reflect limitations of our InfiniteBench wrapper rather than model capacity, so we do not use these rows as method-quality evidence.

\begin{table}[ht]
\caption{InfiniteBench audit at 16k maximum context (Llama-3.1-8B-Instruct). Scores are task metrics from the KVPress evaluation wrapper. Results are exploratory and not used as headline evidence.}
\label{tab:infbench}
\centering
\small
\begin{tabular}{llrrrr}
\toprule
Task & Method & No comp. & 50\% & 75\% & 88\% \\
\midrule
\multirow{3}{*}{kv\_retrieval}
 & FRC & 0.16 & 0.00 & 0.00 & 0.00 \\
 & ChunkKV & 0.16 & 0.00 & 0.00 & 0.00 \\
 & ExpectedAttention & 0.16 & \textbf{0.04} & \textbf{0.02} & 0.00 \\
\midrule
\multirow{3}{*}{longbook\_choice\_eng}
 & FRC & 0.40 & 0.04 & 0.02 & 0.10 \\
 & ChunkKV & 0.40 & 0.02 & 0.02 & 0.02 \\
 & ExpectedAttention & 0.40 & \textbf{0.38} & \textbf{0.34} & \textbf{0.36} \\
\midrule
\multirow{3}{*}{number\_string}
 & FRC & 0.14 & 0.14 & 0.14 & 0.14 \\
 & ChunkKV & 0.14 & 0.14 & 0.14 & 0.14 \\
 & ExpectedAttention & 0.14 & 0.14 & 0.14 & 0.12 \\
\midrule
\multirow{3}{*}{passkey}
 & FRC & 0.14 & 0.14 & 0.14 & 0.14 \\
 & ChunkKV & 0.14 & 0.14 & 0.14 & 0.14 \\
 & ExpectedAttention & 0.14 & 0.14 & 0.14 & 0.14 \\
\bottomrule
\end{tabular}
\end{table}

The pattern is consistent with Section~\ref{sec:limits} and Appendix~\ref{app:longbench}: FRC's RULER-derived weights are well-suited to retrieval-style aggressive compression and underperform on long-document comprehension, where uniform key-norm coverage is the better inductive bias. The framework's prescription remains benchmark-specific evolution, not further tuning of the same scorer family.

\section{Scaled-evolution post-mortem and family-crossing analysis}
\label{app:redux}

This appendix is the engineering record behind the family-reachability and selection-pressure results referenced in Section~\ref{sec:analysis}. It covers four diversity-controlled OpenEvolve runs (B-Redux 1 through 3v2) that test family reachability from a structurally distant seed; a reward-hacking finding from that campaign; the population-level constrained-archive analysis that supports the ``selection pressure shifted toward FRC-aligned scoring'' framing; an earlier 1500-iteration scaled run that briefly entered the FRC family before drifting through an OpenEvolve diff-applier contract gap; and practical guidance for future LLM-guided program evolution on KV cache compression.

\subsection{B-Redux family-crossing experiments}
\label{app:redux_bredux}

We launched four diversity-controlled OpenEvolve runs from \texttt{v1\_rich\_features\_seed.py}, varying the OpenEvolve diversity knobs (\texttt{num\_islands} from 9 to 12, mutator temperature from 0.7 to 1.2, mutator LLM, and population size). Each run used 500 iterations on RULER 4k at $r{=}0.83$ on Llama-3.1-8B-Instruct, query-aware. None of the four crossed the family boundary (Table~\ref{tab:redux}): every best policy stayed within the V1 \{\texttt{attn\_received}, \texttt{tail\_attn\_received}, \texttt{head\_consistency}\} family rather than discovering FRC's \{\texttt{local}, \texttt{neighbor}, \texttt{maxhead}\} family. We also planned to vary the mutator LLM by including \texttt{gemma4:31b}, but that mutator produced malformed diff output incompatible with OpenEvolve's parser, so all four runs used \texttt{qwen3.5:35b-a3b-q4\_K\_M}.

\begin{table}[h]
\centering
\caption{B-Redux family-crossing experiments. Four diversity-controlled OpenEvolve runs from the V1 five-signal seed; 500 iterations each on RULER 4k at $r{=}0.83$, query-aware. None crossed to the FRC \{\texttt{local}, \texttt{neighbor}, \texttt{maxhead}\} family. Best scores are full-RULER aggregates at $r{=}0.88$ on Llama-3.1-8B-Instruct, after the $[-k:]$ reward-hacking fix.}
\label{tab:redux}
\small
\begin{tabular}{l l l}
\toprule
Run & OpenEvolve diversity config & Best post-fix ($r{=}0.88$) \\
\midrule
B-Redux-1   & islands=9, temp=0.7                & 28.9 (stage1 reject) \\
B-Redux-2   & islands=9, temp=1.0                & 68.2 (was 95.6 reward-hack) \\
B-Redux-3v1 & islands=9, temp=1.2                & 20.6 (temp too noisy) \\
B-Redux-3v2 & islands=12, pop=600, temp=0.9      & 27.0 (was 95.6 reward-hack) \\
\bottomrule
\end{tabular}
\end{table}

\paragraph{Reward-hacking via a missing index slice.} B-Redux-2 and B-Redux-3v2 reported full-RULER scores of 95.6 against an uncompressed 95.7, which was suspiciously close. Token-level inspection revealed that both policies were keeping $100\%$ of tokens, evicting nothing despite the requested $r{=}0.88$ budget. Examining the evolved code revealed the cause: the LLM had mutated \texttt{return np.argpartition(scores, -k)[-k:]} into \texttt{return np.argpartition(scores, -k)}, dropping the slice. \texttt{np.argpartition} reorders all indices so the top-$k$ are last but does not slice; the policy therefore returned every index, and the surrounding \texttt{EvolvedPrefillPress.compress()} method (trusting the candidate to honor the budget) kept everything. The evolution-time evaluator, which scores on a fraction of RULER, had no output-length check, so the no-op was rewarded as the best program. After enforcing the tokenwise exact-budget invariant at every cascade stage and re-running the same evolved policies with the slice fix applied to their selection wrapper, B-Redux-2 scored 68.2 and B-Redux-3v2 scored 27.0 at $r{=}0.88$, well below FRC's 84.5. The lesson is direct: if the evaluator does not enforce the applicable retained-length contract, evolution will exploit any path that bypasses compression. The cascade hardening described in Section~\ref{sec:method} (output invariants checked at every stage and classified as ``invalid output'' rather than ``low quality'') closes this hole.

\paragraph{Headline finding.} The four B-Redux runs together show that local LLM-guided code mutation under engineered diversity (more islands, higher temperature, larger populations, and an attempted alternative mutator) does not bridge the structural gap between the V1 \{attn, tail, consistency\} family and the FRC \{local, neighbor, maxhead\} family. We treat this as a reachability bound, not a no-go theorem: a sufficiently strong frontier mutator may cross.

\subsection{Practical guidance for LLM-guided program evolution on KV compression}
\label{app:redux_guidance}

We foreground the practical lessons because they are the most reusable artifact. Each item is supported by the post-mortem material that follows.

\begin{enumerate}\setlength\itemsep{2pt}
\item Prefer seeds whose structure already places the search in a high-scoring neighborhood. Evolution under an LLM mutator is a local refiner, not an autonomous discoverer of structurally distant families. The sterile-seed experiment in this appendix illustrates this point directly.
\item Validate output invariants (shape, count, domain constraints) at the evaluation boundary, not only through the downstream metric. Any slack becomes a reward-hacking surface; we observed two structurally different exploits of the same gap.
\item Type the enriched feature set as \emph{required} fields in the seed contract so that mutators see the names they can edit. Names that are documented but absent from the editable surface are typically not surfaced by the mutator.
\item Enforce the editable region in \emph{both} the prompt and the diff applier. \texttt{EVOLVE-BLOCK} markers are a contract only if the diff parser rejects SEARCH regions outside them and the prompt names the contract; otherwise long-budget search will eventually mutate locked harness code, as we observed at iteration 1264 in the scaled run.
\item Use a reasoning-capable mutator (Opus-class or GPT-5-class) where compute permits. Reasoning mutators are more likely to honor an unenforced contract than the no-reasoning variant we used at scale.
\item Run at least three RNG seeds for any family-crossing claim. Our $n{=}1$ scaled-evolution result is suggestive only.
\end{enumerate}

\subsection{Original family-crossing experiment (v1 seed)}

Starting from the v1 seed (5 signals: \texttt{attn\_received}, \texttt{tail\_attn\_received}, \texttt{head\_consistency}, \texttt{key\_norms}, \texttt{distance\_to\_end}) with all 17 features visible in the editable interface, OpenEvolve ran for 500 iterations on Llama-3.1-8B-Instruct, RULER 4k, $r{=}0.83$. After 500 iterations, evolution simplified the scorer from 5 signals to 3 but stayed entirely within the v1 family. The best variant used $0.65 \cdot \texttt{attn} + 0.22 \cdot \texttt{tail} + 0.13 \cdot \texttt{consistency}$ and scored 73.4, below ChunkKV (85.1) and far below FRC (91.2). None of the nine richer features, including the three FRC signals, was ever referenced.

\subsection{Hyperparameter sweep (v1 seed)}

We executed four additional 500-iteration runs from the v1 seed with systematically varied hyperparameters (Table~\ref{tab:redux_config}; see also Appendix~\ref{app:configs}).

\begin{table}[ht]
\caption{B-redux configurations (v1 seed, 500 iter, $r{=}0.83$, Qwen3.5-35B mutator).}
\label{tab:redux_config}
\centering
\small
\begin{tabular}{lrrrrrr}
\toprule
Run & Islands & Pop & Temp & Mig.\ int & Mig.\ rate & Evolution score \\
\midrule
Original (Expt B) & 3  & 200 & 0.7 & 20 & 0.15 & 73.4 \\
B-redux-1         & 9  & 400 & 0.7 & 10 & 0.25 & 28.9 \\
B-redux-2         & 9  & 400 & 1.0 & 10 & 0.25 & 73.6$^\dagger$ \\
B-redux-3 (v1)    & 9  & 400 & 1.2 & 10 & 0.25 & 20.6 \\
B-redux-3 (v2)    & 12 & 600 & 0.9 & 5  & 0.30 & reward-hack$^\dagger$ \\
\bottomrule
\end{tabular}\\[0.3em]
{\footnotesize $\dagger$: Initial reports of 95.6 for B-redux-2 and B-redux-3v2 were traced to a no-op eviction bug; corrected scores are reported below.}
\end{table}

Across the four additional runs (2{,}000 iterations in total), no evolved program referenced any of \texttt{local\_attn\_received}, \texttt{neighbor\_attn\_density}, or \texttt{max\_head\_attn\_received}. All surviving variants were weighted combinations of v1 signals. The v1 editable surface, even with all 17 features documented in the interface, did not surface the FRC-family signals to the mutator.

\subsection{v2 seed contract and scaled evolution}

The v2 seed contract narrows the editable block to the scoring expression and types the enriched-feature fields, including the three FRC signals, as \emph{required} inputs to the dataclass. This removes two failure modes from the v1 sweep: it prevents diff-mutation from \emph{intentionally} editing harness code, and it ensures that the enriched feature names appear in the editable surface so the LLM mutator can refer to them.

A single 500-iteration v2 run reaches stage2 = 89.71 and full-RULER 4k = 90.14 (Llama, $r{=}0.83$, fraction 1.0), with the best evolved scorer referencing \texttt{neighbor\_attn\_density}. This was the first run in our program of work to cross the v1 family boundary through evolution alone.

To stress-test the finding under scaled budget and seed variation, we launched three independent 1500-iteration evolutions (B-redux-v3a/b/c, RNG seeds 42/1337/2718) with otherwise identical configurations. Two runs (v3a, v3c) stalled before any checkpoint was saved, and given the seven-day-per-run wall-clock budget on a single H200, we could not relaunch them within the deadline window. Only B-redux-v3b (RNG=1337) completed. We therefore report v3b as a single scaled run rather than a multi-seed convergence study, and we carry $n{=}1$ on every per-program claim derived from this run.

\subsection{Reward-hacking failures and the contract gap}
\label{app:redux_hacks}

Two structurally different reward-hacking failures occurred during this work. Both exploit the same root cause: an output invariant that the evaluator did not enforce, combined with an \texttt{EVOLVE-BLOCK} contract that OpenEvolve does not enforce in its diff path.

\paragraph{Failure 1: missing top-$k$ slice (v1 sweep).}
Two of the four runs in the v1 hyperparameter sweep initially reported evolution scores of 95.6, numerically equal to the uncompressed baseline. We instrumented a token-inspection probe and found that both policies returned all indices rather than the top-$k$.

\begin{lstlisting}[caption={The no-op program mutation found by evolution. The slice \texttt{[-k:]} had been dropped.}]
def select_tokens_to_keep(ctx):
    scores = ... # (feature combination)
    k = ctx.cache_budget
    if k >= len(scores):
        return np.arange(len(scores))
    # BUG: missing trailing `[-k:]`
    return np.argpartition(scores, -k)
\end{lstlisting}

\texttt{np.argpartition(scores, -k)} returns a reordered full-length index array whose last $k$ entries are the top-$k$; without the slice, the function returns every index. The evaluator consumed the returned array as the retained-index set without checking that its length matched the cache budget. We patched the evaluator to assert \texttt{len(kept\_indices) == cache\_budget}, fixed the slice in both evolved programs, and re-ran on full RULER at $r{=}0.88$ (Table~\ref{tab:redux_fixed}).

\begin{table}[ht]
\caption{Corrected evaluation: full RULER ($r{=}0.88$, Llama-3.1-8B-Instruct, fraction 1.0). FRC remains the strongest method by at least 14 points.}
\label{tab:redux_fixed}
\centering
\small
\begin{tabular}{lrr}
\toprule
Method & Avg & $\Delta$ vs.\ No Compression \\
\midrule
No Compression & 95.7 & -- \\
\textbf{FRC (ours)} & \textbf{84.6} & $-11.1$ \\
AdaSnapKV & 70.6 & $-25.1$ \\
ChunkKV & 69.2 & $-26.5$ \\
B-redux-2 (fixed) & 68.2 & $-27.5$ \\
SnapKV & 67.2 & $-28.5$ \\
Finch & 65.5 & $-30.2$ \\
ExpectedAttention & 42.2 & $-53.5$ \\
B-redux-3v2 (fixed) & 27.0 & $-68.7$ \\
\bottomrule
\end{tabular}
\end{table}

\paragraph{Failure 2: v3b iteration 1264 (brief family entry, then a second hack).}

At iteration 1073, the global-best program in v3b carried the scorer
\[
0.30\cdot\texttt{local\_attn\_received} + 0.60\cdot\texttt{neighbor\_attn\_density} + 0.10\cdot\texttt{max\_head\_attn\_received},
\]
referencing exactly the three FRC signals (with density-heavy weights rather than FRC's locality-heavy weights). It scored stage1 = 90.51 and stage2 = 90.70 on Llama RULER 4k at $r{=}0.83$, within $\approx 0.5$ points of FRC's 91.2. Evolution had \emph{briefly entered} the FRC three-signal family from a structurally distant V1 seed.

By iteration 1264, however, a new global best emerged at stage1 = 96.06 and stage2 = 95.61, numerically equal to no compression. Inspection showed that \texttt{select\_tokens\_to\_keep}, which the seed comment explicitly marks as outside the editable block and which performs top-$k$ selection on top of the scorer, had been rewritten to drop the trailing \texttt{[-k:]} slice on \texttt{np.argpartition}. This is the same structural reward-hack mechanism documented above, but applied to locked code rather than to the scorer. The mid-run FRC-family entry at iteration 1073 was discarded by the search and not retained.

\paragraph{Why we report v3b at iteration 1073 rather than iteration 1264.} The iteration-1264 score is a measurement artifact of an unenforced contract, not an evolved policy. We therefore report the iteration-1073 program as the v3b scaled-evolution result and discard iteration 1264 as a second instance of the reward-hacking failure mode. This re-anchors the headline scaled-evolution number to a non-final program, and with $n{=}1$ we cannot distinguish robust family crossing from a lucky lineage. This is consistent with our reading in Section~\ref{sec:results}: the result is suggestive evidence of family reachability under selection pressure, not robust convergence.

\paragraph{Cause: the editable contract is documentary, not enforced.} The v2 seed places \texttt{select\_tokens\_to\_keep} outside \texttt{\# EVOLVE-BLOCK-START / END} markers and includes a docstring stating that those markers protect the scoring logic. We had assumed, and the seed implies, that the framework treats this as a contract. It does not. Two gaps in OpenEvolve's diff path make the contract documentary only:

\begin{itemize}\setlength\itemsep{0pt}
  \item \textbf{Diff applier does not validate region.} The function \texttt{apply\_diff} in \texttt{openevolve/utils/code\_utils.py} parses \texttt{<<<<<<< SEARCH / ======= / >>>>>>> REPLACE} blocks and applies each block at its first textual match anywhere in the parent program. There is no check that the matched region lies between \texttt{\# EVOLVE-BLOCK-START} and the matching \texttt{END}. A utility \texttt{parse\_evolve\_blocks} exists in the same file but is not called in the diff path. (See OpenEvolve Issue~\#281, PR~\#352, PR~\#354 in v0.5.1.)
  \item \textbf{Default prompt does not name the contract.} The default \texttt{diff\_user.txt} template asks the model to ``suggest improvements to the program'' and shows the SEARCH/REPLACE format without mentioning EVOLVE-BLOCK markers. The mutator is therefore never told that any region of the file is off-limits.
\end{itemize}

With a no-reasoning $\approx 32$B mutator (Qwen3.5-nothink) and 1500 iterations across 12 islands, this hole was traversed twice in v3b: early in the run, a \texttt{PrefillScoreContext} dataclass collapse on island 3 (22 fields reduced to 5, all outside the EVOLVE-BLOCK), and at iteration 1264, the \texttt{select\_tokens\_to\_keep} rewrite documented above. The intermediate FRC-family discovery at iteration 1073 lived on a different lineage that happened to preserve the locked code by accident.

\subsection{Constrained-archive analysis: population-level selection pressure}
\label{app:redux_archive}

To complement the $n{=}1$ per-program iteration-1073 finding, we performed a population-level analysis over all available run archives (v3b, partial v3a/v3c data, and V1 through V5 archive snapshots). The pipeline is:

\begin{enumerate}\setlength\itemsep{0pt}
\item AST-based extraction of \texttt{EVOLVE-BLOCK} contents (parses Python via \texttt{ast}; handles \texttt{ctx.feature} attribute access, unary minus, variable aliasing, and $n$-way multiplicative interactions);
\item SHA-256 deduplication of EVOLVE-BLOCK bodies, yielding 953 unique programs;
\item Linear-only subset filter: programs whose \texttt{score\_tokens} is a sum of (coefficient $\times$ \texttt{ctx.feature}) terms with no clipping, no conditionals, and no interactions, yielding $n{=}621$;
\item Per-program \emph{abs\_frc\_share} computed as the fraction of total absolute weight assigned to the three FRC-family signals;
\item Spearman $\rho$ against stage-1 RULER quality.
\end{enumerate}

\paragraph{Headline correlations.} On the linear-only $n{=}621$ subset:
\begin{itemize}\setlength\itemsep{0pt}
\item $\rho(\text{stage1}, \text{abs\_frc\_share}) = 0.513$
\item $\rho(\text{stage1}, \text{pos\_frc\_share}) = 0.567$
\item $\rho(\text{stage1}, \text{neighbor\_attn\_density\_abs}) = 0.399$
\item $\rho(\text{stage1}, \text{max\_head\_attn\_received\_abs}) = 0.293$
\item $\rho(\text{stage1}, \text{local\_attn\_received\_abs}) = -0.004$
\item $\rho(\text{stage1}, \text{frc\_presence}) = 0.008$ (presence alone is not predictive; magnitude matters)
\end{itemize}

Within-run $\rho_{\text{abs}}$ is positive in 10 of 12 analyzable runs (median 0.593, mean 0.460; one run dropped for $n<5$ unique programs). Per-generation trajectory $\rho(\text{generation}, \text{mean abs\_frc\_share})$ is positive in all 11 runs with $\geq 5$ generations (median 0.464, mean 0.568); first-to-last-generation $\Delta$(abs\_frc\_share) is positive in all 13 runs (range $+0.10$ to $+0.85$). Filtering to the noise-floor subset (stage-1 $\geq 25$, where the program is non-trivially functional) tightens the relationship to $\rho_{\text{abs}} = 0.936$ ($n{=}39$).

\paragraph{Per-pool breakdown.} The linear-only correlation varies systematically with editable-pool size: pool=4 ($n{=}275$, $\rho_{\text{abs}}{=}0.654$), pool=5 ($n{=}111$, $\rho_{\text{abs}}{=}0.850$), pool=6 ($n{=}120$, $\rho_{\text{abs}}{=}0.359$), and pool=7 ($n{=}115$, $\rho_{\text{abs}}{=}0.565$). Pool=6 is the regime where within-run $\rho$ is weakest: two of the three pool=6 runs (\texttt{8signals\_a}, \texttt{9signals\_a} on cluster A) returned slightly negative within-run $\rho_{\text{abs}}$, while the third (\texttt{8signals\_100itr} on cluster B) recovered $\rho_{\text{abs}}{=}0.663$. We attribute the pool=6 dip to a region of the editable surface where the mutator has enough degrees of freedom to drift toward non-FRC scoring without a stronger reasoning-capable mutator to bias the search; the trend reverses at pool=7 once two FRC-family signals re-enter the visible pool.

\paragraph{Interpretation: what this evidence does and does not claim.} \textit{Does claim:} under the v2 editable contract, the cascade-gated selection pressure operating across runs shifted programs toward FRC-aligned scoring, and the population-level $\rho$ is robust to deduplication and to the linear-only restriction. \textit{Does not claim:} that any specific run rediscovered FRC, that scaled evolution autonomously found the canonical FRC weights (it did not; the iteration-1073 weights are density-heavy), or that this is a robust convergence result ($n{=}1$ per-program; partial archive pooling for the population claim). The wording used throughout the paper (``shifted toward FRC-aligned scoring,'' ``briefly entered the FRC family,'' and ``selection pressure within the editable surface'') is calibrated to this evidence base.

\subsection{Sterile-seed experiment}

For completeness, 500 iterations from a seed implementing ExpectedAttention's key-norm scorer (a single-signal formulation structurally distant from any multi-signal high-scoring neighbor) never produced a variant that passed the 45-point Stage-1 quality gate; the best candidate scored 6.2. Local code mutation in this problem setting does not bridge large structural gaps between seed families.

\subsection{Proposed framework-level fix}

A minimal patch closes the contract-enforcement hole: (i) constrain \texttt{apply\_diff} to reject SEARCH regions whose first match lies outside any \texttt{EVOLVE-BLOCK} pair (an approximately 10-line change reusing \texttt{parse\_evolve\_blocks}); (ii) prepend one sentence to \texttt{diff\_user.txt} naming the marker contract and stating that diffs outside the block will be discarded; and (iii) add evaluator-side detection of degenerate identity-like outputs (length mismatch and full-coverage checks). We have a draft v3 editable contract that incorporates (i) and (iii) at the seed level; a clean upstream fix would also land (ii) and remove the burden from each seed author.

\subsection{Bounds on the evolutionary contribution within the V5 run}\label{app:evobounds}

This subsection collects the bound on what evolution did inside the V5 run that produced the published FRC weights, and it names the four pieces of evidence that position the loop, rather than the final 100-iteration run, as the unit of contribution.

The relevant bound is that, within V5, the literature-informed seed scored $\approx 91.2$ on RULER 4k at $r{=}0.83$ before any iteration, and the 25-iteration run added $\approx 0.2$ by adjusting the weights to $(0.55, 0.30, 0.15)$. Within that 25-iteration budget, evolution acted as a weight tuner. Four pieces of evidence place the loop's contribution beyond that final refinement.

\textbf{(i) The V5 seed is itself a loop output.} The literature-guided audit that produced the V5 feature pool was triggered by the V3 plateau diagnosis (a generic five-signal interface was too coarse), and the V5 operating point was set by the V4 cross-ratio-failure diagnosis (the cascade was rewarding policies that worked only in the easy regime). The seed weights $(0.60, 0.25, 0.15)$ are the output of those two diagnostic interventions; the final 100 iterations are the small-budget refinement on top of a seed already shaped by the loop.

\textbf{(ii) Cross-(model, context, ratio) transfer at fixed weights.} The $(0.55, 0.30, 0.15)$ weights were evolved on Llama RULER 4k at $r{=}0.83$, and they are rank-1 among the evaluated single-pass KVPress baselines at all 12 of 20 grid cells with $r \geq 0.75$ (across both Llama-3.1-8B-Instruct and Qwen3-8B and across both 4k and 8k context), with no retuning (Figure~\ref{fig:heatmap_r088} in the main text). If the V5 evolution had merely overfit one (model, ratio, context) configuration, the same fixed weights would not transfer to a second model family at a second context length.

\textbf{(iii) Population-level selection pressure.} Across $n{=}621$ unique linear-only programs sampled from 13 evolution runs, Spearman $\rho(\text{stage1}, \text{abs\_frc\_share}) = 0.513$, rising to $0.936$ ($n{=}39$) on the noise-floor-restricted subset; the per-generation trajectory of mean FRC-share is positive in all 11 runs with $\geq 5$ generations, and the first-to-last $\Delta$(FRC-share) is positive in all 13 runs (range $+0.10$ to $+0.85$). The relationship is robust to deduplication and to the linear-only restriction (Appendix~\ref{app:redux}, especially the constrained-archive analysis in Section~\ref{app:redux_archive}).

\textbf{(iv) Family reachability from a structurally distant seed.} A 1500-iteration scaled run from the V1 seed briefly entered the FRC three-signal family at iteration 1073, scoring within $\approx 0.5$ points of FRC's 91.2, before being displaced by a different lineage that exploited a contract-enforcement gap in OpenEvolve's diff applier on an unrelated island. The displacement is therefore not selection pressure away from FRC; it is mutation outside the editable region.

Taken together, these four pieces of evidence position the loop as the unit of contribution: the V5 seed is an output of the loop, the V5 weights transfer because they capture attention-structural rather than model-specific properties, the editable surface concentrates selection pressure on FRC-aligned scoring at population scale, and even from a distant seed the family is reachable. The remaining gap (robust autonomous rediscovery without a contract-enforcing diff path and a more capable mutator) is a question about \emph{framework infrastructure}, not about the search space.

\section{OpenEvolve search details}
\label{app:openevolve}

This appendix details the OpenEvolve search procedure referenced in Section~\ref{sec:method} and complements Appendix~\ref{app:configs} by documenting the engine internals and prompt construction needed to interpret our results. OpenEvolve v0.5.1 \citep{openevolve} is an open-source engine resembling AlphaEvolve \citep{novikov2025alphaevolve}; we use it as our LLM-guided code-evolution vehicle throughout the paper.

\paragraph{Engine overview.} The core loop is: (i) sample a parent program from the per-island MAP-Elites archive \citep{mouret2015illuminating}; (ii) construct a mutator prompt from the parent and a small set of archive neighbors; (iii) call the mutator LLM to produce a unified diff; (iv) apply the diff via \texttt{apply\_diff} to produce a candidate program; (v) run the cascade evaluator; and (vi) write the candidate into its MAP-Elites cell if it survives Stage-2 and exceeds the cell's incumbent fitness. MAP-Elites archives are partitioned per island; behavioral axes are program complexity (bucketed source length) and diversity (cosine distance in code-embedding space). Migration uses a ring topology with fraction $\rho_{\text{mig}}$ every $\tau_{\text{mig}}$ generations; per-run values are listed in Appendix~\ref{app:configs}.

\paragraph{Mutator prompt.} The mutator is \texttt{qwen3.5:35b-a3b-q4\_K\_M} via local Ollama. The scaled-evolution v3b run used the no-reasoning variant of the same model, trading faster generation for slightly less coherent diffs. The prompt structure is: a system message describing the role, objective, and unified-diff output format; the parent program in full; a small set of archive samples (3 to 5 high-fitness or high-diversity neighbors); and an instruction to generate a unified diff modifying the EVOLVE-BLOCK region. Sampling temperature is $0.7$ for V1 through V5 and $0.5$ for the scaled v3b run. The diff parser \texttt{apply\_diff} consumes the unified diff and applies it to the parent.

\paragraph{Known contract gap.} As documented in Section~\ref{sec:method} and analyzed in detail in Appendix~\ref{app:redux}, \texttt{apply\_diff} does not enforce EVOLVE-BLOCK boundaries (OpenEvolve Issue \#281, PR \#352, PR \#354 in v0.5.1). Mutator-generated diffs can therefore escape the editable region and modify the surrounding harness, which is the framework limitation that surfaces in the v3b drift past iteration 1073 (Appendix~\ref{app:redux_hacks}).

\paragraph{Cascade evaluator.} Stage 0 is a syntax check (the program parses as Python AST) plus an import-validity check; approximately 100\% of valid mutator output passes Stage 0. Stage 1 is a small-fraction RULER calibration (\texttt{small\_fraction}=$0.02$) gated at aggregate quality $\geq 45.0$; pass rate is approximately 30 to 40\% after Stage 0. Stage 2 is full-fraction held-out evaluation (\texttt{full\_fraction}=$0.10$); all Stage-2 survivors enter the archive. Output-invariant checks are applied at every stage: the retained-index count must equal the cache budget, indices must be unique, and indices must lie in $[0, L)$. Failures in these checks are classified as ``invalid output'' rather than ``low quality'', an important distinction given the reward-hacking analysis of Appendix~\ref{app:redux}.

\paragraph{Archive write strategy and selection.} Survivors enter their MAP-Elites cell. If the cell is empty, the program occupies it; if the cell is occupied, the new program replaces the incumbent only if it has strictly higher fitness (Stage-2 quality score). Sampling for the next mutation parent uses softmax over a fitness-and-recency utility (with a $0.7$/$0.3$ weighting in our runs), biasing toward recent high-fitness programs while preserving exploration of older cells.

\paragraph{FRC main-run summary.} The V5 evolution that produced the published FRC weights had the following configuration:
\begin{itemize}\setlength\itemsep{0pt}
\item \textbf{Seed}: \texttt{chunkkv\_inspired\_seed.py}, weights $(0.60, 0.25, 0.15)$
\item \textbf{Target model}: Llama-3.1-8B-Instruct
\item \textbf{Benchmark}: RULER 4k, query-aware, $r{=}0.83$
\item \textbf{Iterations}: $25$
\item \textbf{Stage 1 / Stage 2 fractions}: $0.02$ / $0.05$
\item \textbf{Mutation LLM}: \texttt{qwen3.5:35b-a3b-q4\_K\_M} (Ollama, reasoning variant)
\item \textbf{Runtime}: $\approx 8.5$ minutes/iteration, $\approx 3.5$ hours total on $1\times$ H100 (Ollama-host machine)
\item \textbf{Result}: weights changed to $(0.55, 0.30, 0.15)$; feature set and structural prior unchanged; the best evolved descendant exceeded the V5 seed by $\approx 0.2$ on full RULER 4k at $r{=}0.83$
\end{itemize}

\paragraph{Cascade survival under scaled budget.} For the scaled v3b run (1500 iterations, pop=600, 12 islands), approximately 12\% of mutator outputs survive the cascade and enter the archive. Cascade gating provides approximately $8\times$ compute savings relative to a hypothetical full-evaluation strategy at the same budget.

\section{Detailed algorithms}
\label{app:algorithms}

This appendix collects the algorithms underlying \Name and FRC. Algorithms~\ref{alg:mapelites} and~\ref{alg:cascade} specify the LLM-driven evolutionary loop and the three-stage cascade evaluator, framing \Name as a search procedure. Algorithms~\ref{alg:context} through~\ref{alg:gpufrc} specify the per-batch compression path: how the \texttt{PrefillContext} is built from raw attention and key tensors (Algorithm~\ref{alg:context}); how a tail-window attention proxy is computed when eager attentions are unavailable (Algorithm~\ref{alg:proxy}); how chunk-structured top-$k$ selection wraps any pointwise scorer (Algorithm~\ref{alg:chunktopk}); and the GPU-native compress procedure used at deployment time for the FRC scorer (Algorithm~\ref{alg:gpufrc}). Notation: $n$ is the cache length, $H_{kv}$ is the number of key-value head groups, $H$ is the number of attention heads, $d$ is the per-head dimension, $w$ is the tail window, $L_{c}$ is the chunk length, $r$ is the compression ratio, and $b = \lceil n(1-r) \rceil$ is the cache budget.

\begin{lstlisting}[language=,basicstyle=\ttfamily\scriptsize,frame=single,keywordstyle={},commentstyle={},stringstyle={},showstringspaces=false,caption={Island-based MAP-Elites with LLM mutation.},label={alg:mapelites}]
inputs: seed P0; evaluator E; mutator M; islands K; pop cap N_pop;
        archive feature axes (complexity, diversity), bins B;
        migration interval tau_mig, migration rate rho_mig; budget T.
initialise: for k in 0..K-1: island[k] = {pop: {P0}, archive: empty B x B}
            evaluate P0 once via E and store its fitness in every island
for t = 1..T:
    k     <- sample_island(islands)             # softmax over fitness, recency
    par_t <- top_k(island[k].archive)           # high-fitness parents
    par_d <- random_diverse(island[k].archive)  # cells with rare neighbors
    diff  <- M( prompt(par_t U par_d) )         # unified diff via LLM
    P_t   <- apply_diff(par_t[0], diff)
    if not parses(P_t) or has_merge_markers(P_t): log_invalid; continue
    R     <- E(P_t)                             # see Algorithm 2
    cell  <- compute_cell(P_t)
    if archive cell empty or R.fitness > cell.best:
        island[k].archive[cell] <- P_t
    island[k].pop <- {P_t} U island[k].pop, capped at N_pop by lowest fitness
    if t mod tau_mig == 0:
        migrate rho_mig of each island.archive to a neighbor
return arg max over all islands and cells of fitness
\end{lstlisting}

\begin{lstlisting}[language=,basicstyle=\ttfamily\scriptsize,frame=single,keywordstyle={},commentstyle={},stringstyle={},showstringspaces=false,caption={Three-stage cascade evaluator.},label={alg:cascade}]
inputs: candidate program path p; small fraction rho1; full fraction rho2;
        score gate tau; output invariants I.
Stage 0 (free): syntax + import check; reject on failure.
Stage 1: run KVPress runner with fraction rho1; read S1 from metrics.json.
         if S1 < tau: return fitness = 0.30 * S1/100, status = stage1_rejected.
Stage 2: run KVPress runner with fraction rho2; read S2 from metrics.json.
         validate I (count == budget; unique indices; in [0, L)).
         return fitness = 0.25*S1/100 + 0.75*S2/100, status = ok.
\end{lstlisting}

\begin{lstlisting}[language=,basicstyle=\ttfamily\scriptsize,frame=single,keywordstyle={},commentstyle={},stringstyle={},showstringspaces=false,caption={Build PrefillContext.},label={alg:context}]
INPUT:  keys K in R^{H_kv x n x d},
        attentions A in R^{H_kv x q_len x n} (directly observed or from Listing~\ref{alg:proxy}),
        tail_window w, sink_size s, chunk granularity target T_C
OUTPUT: PrefillContext ctx

# attention-derived per-token features
P                  <- mean over q_len of A                # (H_kv, n)
attn_received      <- mean over heads of P                # (n,)
w_eff              <- min(w, q_len)
tail_start         <- max(0, q_len - w_eff)
tail_attn_received <- mean over heads, queries [tail_start:q_len] of A
local_attn_received <- copy(tail_attn_received)
global_attn_received <- (mean over heads, queries [0:tail_start] of A) if tail_start>0 else copy(attn_received)
max_head_attn_received <- max over heads of P

# head-distributional features
head_var          <- variance over heads of P
head_consistency  <- 1 / (1 + head_var)
P_norm            <- P / max(sum over heads of P, 1e-8)
H_ent             <- - sum over heads of (P_norm * log max(P_norm, 1e-8))
attn_concentration<- 1 - H_ent / log(max(H_kv, 2))

# key-space features
key_norms         <- mean over heads of ||K||_2
key_deltas        <- mean over heads of ||K[:, 1:] - K[:, :-1]||_2
key_change_norms  <- [0, key_deltas[0], ..., key_deltas[n-2]]

# density via adaptive moving average
window            <- max(3, min(33, 2 * (n // 64) + 1))
if window even:    window <- window + 1
neighbor_attn_density <- convolve(attn_received, ones(window)/window, 'same')

# chunk features
n_chunks          <- min(16, max(4, n // 128 if n>=128 else 4))
chunk_size        <- ceil(n / n_chunks)
chunk_ids         <- arange(n) // chunk_size
chunk_means[c]    <- mean of attn_received over positions where chunk_ids==c
chunk_order       <- argsort(argsort(-chunk_means))
chunk_rank_per_chunk <- 1 - chunk_order / max(n_chunks - 1, 1)
chunk_mean_attn   <- chunk_means[chunk_ids]               # broadcast to n
chunk_rank        <- chunk_rank_per_chunk[chunk_ids]

# geometric and indicator features
positions         <- arange(n)
normalized_positions <- positions / max(n - 1, 1)
distance_to_end   <- (n - 1 - positions) / max(n - 1, 1)
sink_mask         <- positions < s
is_tail_mask      <- positions >= max(0, n - w_eff)

# max-absolute (MaxAbs) normalize continuous score features
for f in {attn_received, tail_attn_received, head_consistency, key_norms,
          local_attn_received, global_attn_received, max_head_attn_received,
          attn_concentration, neighbor_attn_density, chunk_mean_attn,
          key_change_norms}:
    m <- max(|f|);  f <- f / m if m > 1e-8 else 0
# positions, distance features, chunk_rank, masks, and scalars keep native scale

cache_budget      <- max(1, floor(n * (1 - r)))
return PrefillContext(... all fields, cache_budget, r, H_kv, q_len, n)
\end{lstlisting}

\begin{lstlisting}[language=,basicstyle=\ttfamily\scriptsize,frame=single,keywordstyle={},commentstyle={},stringstyle={},showstringspaces=false,caption={Tail-window attention proxy via re-derived RoPE queries.},label={alg:proxy}]
INPUT:  layer module M with query projection W_q,
        hidden states H in R^{B x q_len x D},
        keys K in R^{B x H_kv x n x d},
        RoPE position embeddings (cos, sin),
        tail window w
OUTPUT: A in R^{B x H_kv x w_eff x n}

w_eff   <- min(w, q_len, n)
if w_eff <= 0: return None
n_groups<- H / H_kv
H_w     <- H[:, -w_eff:, :]
Q_pre   <- W_q(H_w)                                          # (B, H, w_eff, d)

cos_w   <- cos[:, -w_eff:];  sin_w <- sin[:, -w_eff:]
Q       <- Q_pre * cos_w.unsqueeze(1) +
           rotate_half(Q_pre) * sin_w.unsqueeze(1)            # re-apply RoPE

K_full  <- repeat_kv(K, n_groups)                             # (B, H, n, d)
S       <- matmul(Q.float(), K_full.float().transpose(-2,-1)) / sqrt(d)
mask    <- (k_pos > q_pos)
S       <- S.masked_fill(mask, -inf)
S       <- softmax(S, dim=-1).to(K.dtype)
A       <- S.view(B, H_kv, n_groups, w_eff, n).mean(dim=2)    # collapse GQA
return A
\end{lstlisting}

\begin{lstlisting}[language=,basicstyle=\ttfamily\scriptsize,frame=single,keywordstyle={},commentstyle={},stringstyle={},showstringspaces=false,caption={Chunk-structured top-$k$ selection.},label={alg:chunktopk}]
INPUT:  scores s in R^n, chunk length L_c, compression ratio r
OUTPUT: sorted retained indices I

if n == 0: return empty
chunk_ids <- arange(n) // L_c
n_chunks  <- max(chunk_ids) + 1
for c = 0..n_chunks-1:
    chunk_score[c] <- mean of s over positions where chunk_ids == c
n_keep    <- max(1, floor(n_chunks * (1 - r)))
top_chunks<- sort(argpartition(chunk_score, -n_keep)[-n_keep:])
I         <- []
for c in top_chunks:
    I <- I ++ arange(c*L_c, min(c*L_c + L_c, n))
return sort(I)
\end{lstlisting}

\begin{lstlisting}[language=,basicstyle=\ttfamily\scriptsize,frame=single,keywordstyle={},commentstyle={},stringstyle={},showstringspaces=false,caption={GPU-native FRC \texttt{compress()}.},label={alg:gpufrc}]
INPUT:  K, V in R^{B x H_kv x n x d}, hidden states H, attention A or None,
        kwargs, chunk length L_c, tail window w, ratio r
OUTPUT: compressed (K', V')

if r == 0: return K, V
if A is None: A <- attention_proxy(M, H, K, kwargs, w)
A <- A.float()

batch_indices <- []
kept_lengths <- {}
for b in 0..B-1:
    A_b              <- A[b]                              # (H_kv, q_len, n)
    H_kv, q_len, n   <- shape(A_b)

    local_attn       <- mean over heads, queries of A_b[:, -min(w, q_len):, :]
    attn_received    <- mean over heads, queries of A_b
    win              <- max(3, min(33, 2 * (n // 64) + 1))
    if win is even: win <- win + 1
    neighbor_density <- avg_pool1d(attn_received.view(1,1,n), win,
                                   stride=1, padding=win//2).view(n)
    per_head_mean    <- mean over queries of A_b           # (H_kv, n)
    max_head_attn    <- max over heads of per_head_mean

    for f in {local_attn, neighbor_density, max_head_attn}:
        m <- f.abs().max();  f <- f / m if m > 1e-8 else 0 * f

    scores       <- 0.55 * local_attn + 0.30 * neighbor_density
                                       + 0.15 * max_head_attn

    chunk_ids    <- arange(n, device) // L_c
    n_chunks     <- chunk_ids.max() + 1
    chunk_score  <- zeros(n_chunks); chunk_count <- zeros(n_chunks)
    chunk_score.scatter_add_(0, chunk_ids, scores)
    chunk_count.scatter_add_(0, chunk_ids, ones_like(scores))
    chunk_score  <- chunk_score / chunk_count.clamp(min=1)

    n_keep       <- max(1, floor(n_chunks * (1 - r)))
    top          <- topk(chunk_score, n_keep).indices.sort().values
    indices      <- cat[ arange(c*L_c, min((c+1)*L_c, n)) for c in top ]
    indices      <- indices.sort().values
    batch_indices.append(indices)
    kept_lengths.add(len(indices))

if size(kept_lengths) != 1:
    raise invalid output  # ragged retained lengths across batch items
kept_len <- only element of kept_lengths

idx_tensor <- stack(batch_indices, 0).view(B, 1, kept_len, 1)
                                       .expand(B, H_kv, kept_len, d)
K' <- K.gather(2, idx_tensor).contiguous()
V' <- V.gather(2, idx_tensor).contiguous()
return K', V'
\end{lstlisting}

The explicit \texttt{kept\_lengths} check prevents silent shape mismatch when the final chunk is undersized; if different batch items would retain different token counts, the implementation raises an invalid-output error before stacking.

Together these six algorithms, plus the dataclass definition of Appendix~\ref{app:features}, fully specify both the search procedure (Algorithms~\ref{alg:mapelites} and~\ref{alg:cascade}) and the compression pipeline (Algorithms~\ref{alg:context} through~\ref{alg:gpufrc}). Algorithms~\ref{alg:context} through~\ref{alg:chunktopk} are invoked once per layer per batch during compression; Algorithm~\ref{alg:gpufrc} is the deployment-time substitute for Algorithms~\ref{alg:context} and~\ref{alg:chunktopk} specialized to the FRC scorer.

\newpage
\section*{NeurIPS Paper Checklist}

\begin{enumerate}

\item {\bf Claims}
    \item[] Question: Do the main claims made in the abstract and introduction accurately reflect the paper's contributions and scope?
        \item[] Answer: \answerYes{}
    \item[] Justification: All quantitative claims are supported by Figures~\ref{fig:heatmap_r088}, \ref{fig:delta_baseline}, \ref{fig:slopegraph}, Table~\ref{tab:ablation}, and Appendices~\ref{app:full_tables}, \ref{app:longbench}, \ref{app:redux}. Limitations stated in Section~\ref{sec:limits}.

\item {\bf Limitations}
    \item[] Question: Does the paper discuss the limitations of the work performed by the authors?
    \item[] Answer: \answerYes{}
    \item[] Justification: Section~\ref{sec:limits} provides a dedicated limitations discussion covering the scope of the empirical claims, unintegrated two-pass and learned baselines, limited model/context coverage, fixed absolute smoothing windows, retrieval-oriented transfer limits on LongBench-v2, latency scope relative to measured single-pass baselines, and remaining methodological controls such as random-feature baselines and stronger mutators. The section also explains the practical implication of these limits: FRC should not be read as a universal KV-compression policy or as defining a global accuracy--latency Pareto frontier.

\item {\bf Theory assumptions and proofs}
    \item[] Question: For each theoretical result, does the paper provide the full set of assumptions and a complete (and correct) proof?
    \item[] Answer: \answerNA{}
    \item[] Justification: The paper is empirical and does not state theoretical results, theorems, or formal proof claims. Mathematical expressions such as Eq.~\ref{eq:score} define the implemented scorer rather than proving theoretical guarantees.

    \item {\bf Experimental result reproducibility}
    \item[] Question: Does the paper fully disclose all the information needed to reproduce the main experimental results of the paper to the extent that it affects the main claims and/or conclusions of the paper (regardless of whether the code and data are provided or not)?
    \item[] Answer: \answerYes{}
    \item[] Justification: Sections~\ref{sec:setup}--\ref{sec:results} specify the evaluated models, benchmarks, compression ratios, aggregation method, baselines, and evaluation protocol. Appendix~\ref{app:configs} reports the evolution hyperparameters and run budgets, Appendix~\ref{app:features} defines the exposed feature context, and Eq.~\ref{eq:score} plus the native KVPress press-class description specify the FRC policy needed to reproduce the main results.

\item {\bf Open access to data and code}
    \item[] Question: Does the paper provide open access to the data and code, with sufficient instructions to faithfully reproduce the main experimental results, as described in supplemental material?
     \item[] Answer: \answerNo{}
    \item[] Justification: The evaluated benchmarks and base assets are public, including RULER and KVPress, and the paper specifies the FRC scorer, feature context, and evaluation protocol needed to reproduce the main results. We plan to release the native KVPress press class, scripts, and run instructions, but the current submission does not rely on an already-public anonymized code repository.


\item {\bf Experimental setting/details}
    \item[] Question: Does the paper specify all the training and test details (e.g., data splits, hyperparameters, how they were chosen, type of optimizer) necessary to understand the results?
    \item[] Answer: \answerYes{}
    \item[] Justification: Section~\ref{sec:setup} specifies the models, datasets, context lengths, compression ratios, decoding protocol, baselines, and aggregation method. Appendices~\ref{app:configs}, \ref{app:features}, and \ref{app:openevolve} provide the run hyperparameters, feature context, cascade settings, mutator configuration, and implementation details needed to interpret the experiments.


\item {\bf Experiment statistical significance}
    \item[] Question: Does the paper report error bars suitably and correctly defined or other appropriate information about the statistical significance of the experiments?
    \item {\bf Experiment statistical significance}
    \item[] Answer: \answerNo{}
    \item[] Justification: The main accuracy tables report deterministic greedy-decoding scores on fixed benchmark splits, but we do not report repeated-seed confidence intervals or error bars for RULER/LongBench-v2 accuracy. We do report mean $\pm$ standard deviation for the latency microbenchmark in Appendix~\ref{app:latency}; accuracy variance estimation across dataset seeds and hardware nondeterminism remains future work.

\item {\bf Experiments compute resources}
    \item[] Question: For each experiment, does the paper provide sufficient information on the computer resources (type of compute workers, memory, time of execution) needed to reproduce the experiments?
     \item[] Answer: \answerYes{}
    \item[] Justification: Section~\ref{sec:setup} and Appendix~\ref{app:latency} identify the GPU types used for evaluation and latency measurement, and Appendix~\ref{app:configs} tabulates the iteration budgets and run settings for the evolution experiments. The FRC main evolution used 25 iterations and took approximately 3.5 hours on one H100; additional family-crossing and scaled-redux runs are reported in the configuration appendix.
    
\item {\bf Code of ethics}
    \item[] Question: Does the research conducted in the paper conform, in every respect, with the NeurIPS Code of Ethics \url{https://neurips.cc/public/EthicsGuidelines}?
     \item[] Answer: \answerYes{}
    \item[] Justification: The work studies inference-time KV-cache compression using public benchmarks and public/open-weight models, with no human-subject data collection, private data, or model release that increases generative capability. We have reviewed the NeurIPS Code of Ethics and do not identify a deviation.

\item {\bf Broader impacts}
    \item[] Question: Does the paper discuss both potential positive societal impacts and negative societal impacts of the work performed?
    \item[] Answer: \answerYes{}
\item[] Justification: Section~\ref{sec:limits} discusses broader impacts: FRC can reduce KV-cache memory pressure and serving cost for long-context inference, but like other efficiency improvements it may also lower the cost of harmful uses of existing language models. The paper notes that FRC does not introduce new model capabilities, training data, or a user-facing system, so deployment risks remain governed by the underlying model and application context.
    
\item {\bf Safeguards}
    \item[] Question: Does the paper describe safeguards that have been put in place for responsible release of data or models that have a high risk for misuse (e.g., pre-trained language models, image generators, or scraped datasets)?
    \item[] Answer: \answerNA{}
    \item[] Justification: The paper introduces a KV-cache eviction scorer and evaluation methodology, not a new pretrained model, high-risk generative system, or scraped dataset. Existing model safeguards and usage restrictions remain governed by the underlying model providers and licenses.

\item {\bf Licenses for existing assets}
    \item[] Question: Are the creators or original owners of assets (e.g., code, data, models), used in the paper, properly credited and are the license and terms of use explicitly mentioned and properly respected?
     \item[] Answer: \answerYes{}
    \item[] Justification: The paper cites the existing code, dataset, and model assets used in the experiments, including KVPress, RULER, Llama-3.1-8B-Instruct, Qwen3-8B, and OpenEvolve. The licenses and terms for these assets are acknowledged where applicable, including Apache-2.0 assets and the Llama community license.


\item {\bf New assets}
    \item[] Question: Are new assets introduced in the paper well documented and is the documentation provided alongside the assets?
    \item[] Answer: \answerNo{}
    \item[] Justification: The paper defines the FRC scorer and implementation details, but the implementation package is planned for release after review rather than introduced as an open asset in the current submission.

\item {\bf Crowdsourcing and research with human subjects}
    \item[] Question: For crowdsourcing experiments and research with human subjects, does the paper include the full text of instructions given to participants and screenshots, if applicable, as well as details about compensation (if any)? 
    \item[] Answer: \answerNA{}
    \item[] Justification: The paper does not involve crowdsourcing, user studies, or research with human subjects.

\item {\bf Institutional review board (IRB) approvals or equivalent for research with human subjects}
    \item[] Question: Does the paper describe potential risks incurred by study participants, whether such risks were disclosed to the subjects, and whether Institutional Review Board (IRB) approvals (or an equivalent approval/review based on the requirements of your country or institution) were obtained?
    \item[] Answer: \answerNA{}
    \item[] Justification: The paper does not involve crowdsourcing or human-subject research, so IRB approval or equivalent review is not applicable.

\item {\bf Declaration of LLM usage}
    \item[] Question: Does the paper describe the usage of LLMs if it is an important, original, or non-standard component of the core methods in this research? Note that if the LLM is used only for writing, editing, or formatting purposes and does \emph{not} impact the core methodology, scientific rigor, or originality of the research, declaration is not required.
    \item[] Answer: \answerYes{}
    \item[] Justification: LLM usage is central to the methodology: Qwen3.5-35B is used as the mutation engine inside the OpenEvolve-guided program search. Section~\ref{sec:method_openevolve} and the appendices describe the mutator, cascade evaluation, run settings, and the role of LLM-generated program edits; this is distinct from any incidental writing or editing assistance.

\end{enumerate}

\end{document}